\documentclass[a4paper, 11pt]{article}
\usepackage{my_jcap} % for details on the use of the package, please see the JINST-author-manual

\usepackage{natbib}
\setcitestyle{aysep={}}
\usepackage{graphicx}	% Including figure files
\usepackage{amsmath}	% Advanced maths commands
\usepackage{breqn}

\usepackage{fontawesome5}
\usepackage{color}
\usepackage{xcolor}

\allowdisplaybreaks  % allow page breaks inside align

\usepackage{mdframed}
\newmdenv[
    linecolor=gray,            % Border color
    backgroundcolor=gray!10,    % Background color (20% gray)
    linewidth=0pt,            % Border width
    roundcorner=0pt,            % Rounded corners
    skipabove=\topsep,          % Space above the box
    skipbelow=\topsep,          % Space below the box
    innertopmargin=-5pt,         % Padding inside the box
    innerbottommargin=10pt,
    innerleftmargin=5pt,
    innerrightmargin=5pt,
    splitbottomskip=0pt,        % Space between split parts
    splittopskip=30pt,           % Space between split parts
    nobreak=false,               % Allow breaking across pages
]{mybox}

\newcommand{\kms}{{\rm \, km~s}\ensuremath{^{-1}}}

\newcommand{\msun}{\ensuremath{\, {\rm M}_\odot}} 
         
\newcommand{\mpc}{\ensuremath{\, {\rm Mpc}}}         
\newcommand{\gpc}{\ensuremath{\, {\rm Gpc}}}
     
\newcommand{\hinvmpc}{\ensuremath{\, h{\rm /Mpc}}}

\newcommand{\eg}{{\sl e.g.},\hskip 1pt}

\newcommand*{\vcenteredhbox}[1]{\begingroup
\setbox0=\hbox{#1}\parbox{\wd0}{\box0}\endgroup}

\newcommand{\OrcidID}[1]{ \href[urlcolor = red]{https://orcid.org/#1}{\textcolor{lightgray}{\faOrcid}}}
\newcommand{\OrcidIDName}[2]{\href{https://orcid.org/#1}{#2}}

\newcommand{\Rtwohc}{R_{\rm 200c}}
\newcommand{\Mtwohc}{M_{\rm 200c}}

\newcommand{\fNL}{f_{\rm NL}}
\newcommand{\wpk}{w_{\rm pk}}
\newcommand{\Apk}{A_{\rm pk}}

\newcommand{\kOne}{k_1}
\newcommand{\kTwo}{k_2}
\newcommand{\kThree}{k_3}

\defcitealias{Scoccimarro2012PNGs}{S12}
\defcitealias{Sohn:2023:CMBbest}{S23}
\defcitealias{Sohn:2024:Colliders}{S24}
\defcitealias{paper1}{\textsc{Paper I}}
\defcitealias{paper2}{\textsc{Paper II}}
\defcitealias{paper3}{\textsc{Paper III}}
\usepackage[T1]{fontenc}   % modern font encoding
\usepackage{lmodern}       % Latin-Modern: vectorised version of CM
\usepackage{anyfontsize}   % let NFSS scale to *any* size on demand
\title{\fontsize{19.5pt}{24pt}\selectfont Primordial Physics in the Nonlinear Universe:
signatures of inflationary resonances, \\excitations, and scale dependence}

\author[1, 2, 3]{\OrcidIDName{0000-0003-3312-909X}{Dhayaa Anbajagane}
(\vcenteredhbox{\includegraphics[height=1.2\fontcharht\font`\B]{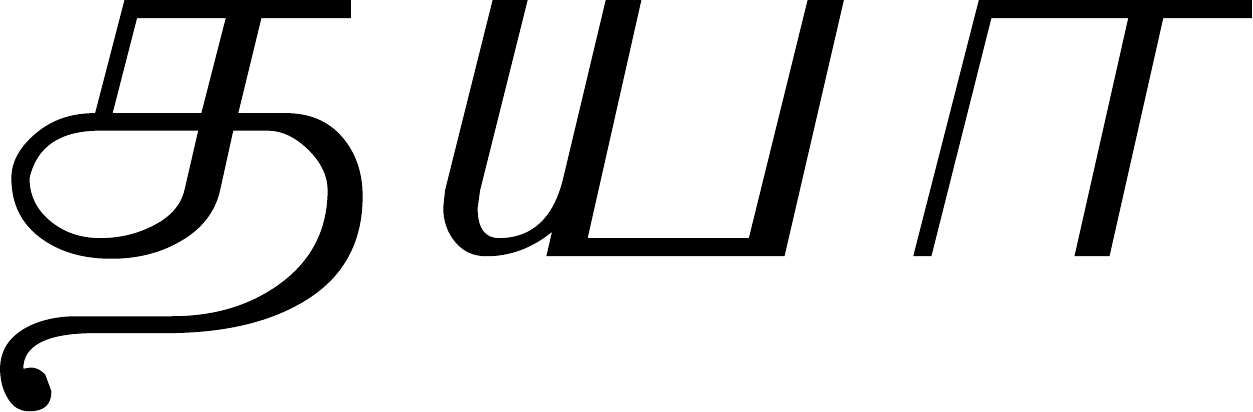}})}
\author[2, 4]{and \OrcidIDName{0000-0002-7577-0806}{Hayden Lee}}

\affiliation[1]{Department of Astronomy and Astrophysics, University of Chicago, Chicago, IL 60637, USA}
\affiliation[2]{Kavli Institute for Cosmological Physics, University of Chicago, Chicago, IL 60637, USA}
\affiliation[3]{NSF-Simons AI Institute for the Sky (SkAI), 172 E. Chestnut St., Chicago, IL 60611, USA}
\affiliation[4]{Center for Particle Cosmology, Department of Physics and Astronomy, University of Pennsylvania, Philadelphia, PA 19104, USA}

\emailAdd{dhayaa@uchicago.edu}
\emailAdd{haydenhl@sas.upenn.edu}

\abstract{Primordial non-Gaussianities (PNGs) are imprints in the initial density field sourced by the dynamics of inflation. These dynamics can induce scale dependence, oscillations, and other features in the primordial bispectrum.
We analyze a suite of over thirty PNG templates, including those used in the \textit{Planck} analyses of the Cosmic Microwave Background (CMB), and resolve their signatures in the deeply nonlinear regime of the late-time density field. 
Using simulations, we forecast results from a lensing analysis of the Year-10 data from the Rubin Observatory Legacy Survey of Space and Time (LSST). 
We find that lensing achieves sensitivity comparable to the CMB for many models, and even surpasses it for templates whose features peak on smaller scales, $k \gtrsim 0.2 \hinvmpc$. 
Many templates generate non-monotonic behaviors in mass and length scales, providing a distinct phenomenology in the resulting late-time structure. 
We simulate, for the first time, resonant signatures consistently in both the primordial power spectrum and bispectrum. 
The constraints on their amplitudes $(\Apk, \fNL)$ are essentially independent, as each affects structure formation in distinct ways. 
Overall, we find that lensing data can provide competitive and complementary constraints on these models, and can deliver leading constraints when the primordial features are predominantly on smaller scales. 
The data products are publicly released as part of the \textsc{Ulagam} simulation suite. Our initial conditions generator is publicly available at \url{https://github.com/DhayaaAnbajagane/Aarambam}.}

\makeatletter
\def\@fpheader{\ }
\makeatother

\begin{document}

\maketitle
\flushbottom

%%%%%%%%%%%%%%%%%%%%%%%%%%%%%%%%%%%%%%%%%%%%%%%%%%

%%%%%%%%%%%%%%%%% BODY OF PAPER %%%%%%%%%%%%%%%%%%

\section{Introduction}

Under the inflationary paradigm, the Universe underwent a rapid phase of expansion in its earlier moments~\citep{Guth:1981:Inflation, Linde:1982:Inflation, Guth2004Inflation}. This process also generates the initial fluctuations in the density field that subsequently grow and are observed in the cosmic microwave background and the large-scale structure. 
The fluctuations are commonly modeled to follow the statistics of a Gaussian random field. 
However, deviations from Gaussianity---called ``Primordial Non-Gaussianities'' (PNGs)---can arise from particle interactions during inflation~\citep[see][for a review]{Chen2010PNGReview}. 
Detecting such PNGs would provide a unique window into the microphysics of the early Universe, offering direct clues about the field content and interactions operative during inflation.

Searches for PNGs have thus far mostly focused on three-point correlations---or ``bispectrum'' in Fourier space---in the CMB \citep{Planck:2014:PNGs, Planck:2016:PNGs, Planck2020PNGs, Sohn:2024:Colliders}, as well as on two-point and three-point correlations in galaxy surveys \citep{Cabass2022MultifieldBOSS, Cabass2022SingleFieldBOSS, Damico2022BossPNG, Philcox2022BossPNG}. 
Notably, these studies are restricted to the (quasi-)linear regime of the density field due to modeling limitations in accessing the non-perturbative regime. 
Consequently, the impact of such signatures on the nonlinear regime of the density field, which is dominated by collapsed structures (i.e., halos) and their dynamics, remains largely unexplored. 
A few works have focused on this regime \citep[][]{Coulton2022QuijotePNG, Jung2023fNLHMFQuijote, Anbajagane2023Inflation}, considering the three standard PNG templates: local, equilateral, and orthogonal types. 

There are also extensions to these standard scenarios that capture interactions of the inflaton with additional massless/massive particles. This framework, known as ``cosmological collider physics'' \citep{Chen:2010:QSF,Baumann:2011nk,Nima:2015:Colliders,Lee:2016vti,Arkani-Hamed:2018kmz}, has recently been studied in the large-scale structure context by~\citet{Goldstein:2024:CosmoColl} for the squeezed-limit bispectrum and by \citet[][henceforth, \citetalias{paper1}]{paper1} for the full bispectrum. 
In particular, in \citetalias{paper1} we introduced a method for generating initial density fields with arbitrary primordial bispectra. This enabled us to resolve, for the first time, over thirty collider PNG templates in the nonlinear regime. 
The inflationary signatures in structure formation from this regime are distinct from those signatures found on linear and quasi-linear scales, and therefore provide a complementary avenue for placing constraints on inflationary models \citep[][\citetalias{paper1}]{Anbajagane2023Inflation}. 
Furthermore, \citetalias{paper1} showed that the potential lensing-only constraints on such collider models can be competitive and complementary to the existing constraints from \textit{Planck} 2018 \citep{Sohn:2024:Colliders}.

Cosmological collider models, however, are only a subset of the wider class of theoretical models that are of interest to the inflation community. For example, analyses of the \textit{Planck} CMB data consider a broad class of bispectra probing a variety of inflationary physics \citep{Planck:2014:PNGs, Planck:2016:PNGs, Planck2020PNGs}. In this work, we apply the methodology of \citetalias{paper1} to study over thirty templates, all of which were considered in these \textit{Planck} analyses. These templates broadly probe resonant particle production, excited initial states, scale dependence etc. This work presents the first study of nonlinear signatures for these bispectrum templates. 
We study their impact on the late-time density field, including the matter power spectra and bispectra, the halo mass function, and the halo bias. We forecast the sensitivity of weak-lensing measurements to such models and compare them to constraints from the CMB. We also consider scenarios where the resonant particle production impacts the power spectrum and bispectrum, and extract the same for this case as well. Our method and the associated simulations are made public.

We organize this paper as follows: Section~\ref{sec:sims} briefly describes our method for incorporating bispectrum signatures into the simulation initial conditions, and the models we consider in this work. Section~\ref{sec:results} presents the signatures of PNGs in the matter field and halo field, and forecasts the constraints obtained from upcoming weak-lensing observations. We summarize in Section~\ref{sec:conclusions}. Appendix \ref{appx:PsecBspecExtra} presents additional results from resonant signatures in both power spectrum and bispectrum, while Appendix \ref{appx:Validation} details a full validation of our initial conditions-generation method for all PNG models considered in this work.

\section{Simulations}\label{sec:sims}

Our simulations are part of the \textsc{Ulagam} suite \citep{Anbajagane2023Inflation}, which contain full-sky lightcones built using the \textsc{PkdGrav3} $N$-body code \citep{Potter2017Pkdgrav3}. The $N$-body simulations follow the same prescription as \citetalias{paper1}: we use $512^3$ particles in a volume of $V = L^3 = (1 \gpc/h)^3$ where $h = H_0 / (100 \kms/\mpc)$ is the dimensionless Hubble constant. Each simulation produces 100 snapshots from $z = 127$ to $z = 0$. The \textsc{PkdGrav3} code automatically produces lightcone shells of the density field for each snapshot. We also run the \textsc{Rockstar} halo finder \citep{Behroozi2013Rockstar} on the particle snapshots to generate halo catalogs. We use the halo mass/radius definition $\Mtwohc = 200 \rho_c \times 4\pi \Rtwohc^3$, where $\rho_c$ is the critical density at a given epoch.

Our procedure for generating initial conditions is detailed in \citetalias{paper1} (see their Section 2). We briefly summarize this method below in Section \ref{sec:sims:ICs}. Then, Section \ref{sec:sims:Models} lists the models considered in this work.

\subsection{Generating Initial Conditions from Arbitrary Bispectra}\label{sec:sims:ICs}

The primordial bispectrum describes three-point correlations of curvature perturbations in Fourier space, and can be expressed as
\begin{equation}
    B(\kOne, \kTwo, \kThree) = 2\fNL K_{12}(\kOne, \kTwo, \kThree) P(\kOne) P(\kTwo) + 2 {\rm \,\, perm.}\,,
\end{equation}
where the form of the kernel $K_{12}$ depends on the inflationary model and $P(k)$ is the primordial power spectrum.  
In this work, we use the kernels presented in the various analyses of the \textit{Planck} CMB data \citep{Planck:2014:PNGs, Planck:2016:PNGs, Planck2020PNGs}.

To generate initial conditions with a given bispectrum, we follow the approach of \citet{Scoccimarro2012PNGs}. This method requires the bispectrum (or kernel) to be separable, i.e., $B(\kOne, \kTwo, \kThree) = f(\kOne)g(\kTwo)h(\kThree)$. This condition is not always met for all bispectra considered in this work. Instead, we follow the approach of \citetalias{paper1} --- which builds on the longstanding mode decomposition efforts in the CMB community \citep[]{Komatsu:2005:KSW, Kenrick:2011:Basis, Fergusson:2012:Modal, Sohn:2023:CMBbest, Sohn:2024:Colliders} and more recently in the large-scale structure community~\citep[\eg][]{Assassi:2017:CLs, Lee:2020:CLs, Chen:2021vba} --- by decomposing the bispectra into a sum of terms that are each individually factorizable as
\begin{equation}\label{eqn:Decompose}
    B(\kOne, \kTwo, \kThree) = \frac{1}{\kOne^2\kTwo^2\kThree^2}\sum_{i,j,k = 0}^{N} \alpha_{ijk} \,q_i(\kOne) q_j(\kTwo) q_k(\kThree)\,.
\end{equation}
The basis functions, $q_i(k)$, are a combination of monomial terms that are supplemented by a series of modified Legendre polynomials,
\begin{align}\label{eqn:modefunc}
    q_i(k) = 
    \begin{cases}
        k^{\frac{4 - n_s}{3} (i - 3)} & \text{if } i \leq 3\,,\\
        \mathcal{P}_{i - 1}(\Tilde{k}) - A_{i - 1} & \text{if } i > 3\,,\\
    \end{cases}
\end{align}
where $\Tilde{k}$ is the normalized wavenumber defined by
\begin{equation}\label{eqn:kbar}
    \Tilde{k}  = -1 + 2\frac{k^x - k_{\rm min}^x}{k_{\rm max}^x - k_{\rm min}^x}\,,
\end{equation}
with $x = (4 - n_s)/3$.
The exponent $x$ accounts for the mild scale dependence of the primordial potential power spectrum. The constant $A_{i - 1}$ term is subtracted to avoid the infrared (IR) divergences in the one-loop power spectrum; see Section 2.1 of \citetalias{paper1} for more details.

Each term in the RHS of Eq.~\eqref{eqn:Decompose} can be used in the method of \citet{Scoccimarro2012PNGs} to generate the primordial potential that sources the late-time density perturbations. 
We can generate $N$ potential fields for the $N$ terms in the summation, one per term. 
The summation of the individual terms then results in a potential that contains the correct bispectrum signal. 
Similar to \citetalias{paper1}, we explicitly validate the numerical accuracy of our method in Appendix \ref{appx:Validation}.

We follow \citetalias{paper1} in using $N = 15$ functions ($q_i$) for our analyses, which results in 680 possible terms in the sum of Eq.~\eqref{eqn:Decompose}. 
In practice, we perform a guided subsampling, as detailed in Appendix A of \citetalias{paper1}, to create a basis set that minimizes correlations between the individual basis terms. Accordingly, we do not use all 680 mode functions and instead use up to 150 modes. 
Finally, we reiterate that our approach to generating initial conditions exactly follows that of \citetalias{paper1}. The only difference in this work is we target a different set of bispectrum templates.

\subsection{Inflationary Models} \label{sec:sims:Models}

We consider a wide variety of models following those explored in \citet{Planck:2014:PNGs, Planck:2016:PNGs, Planck2020PNGs}, and choose a subset to simulate in our work. Many of these models, such as the scale-dependent ones or linear resonances, are manifestly separable \citep{Munchmeyer:2014:OscKSW}. Others that are not manifestly separable due to a factor of $1/(\kOne + \kTwo + \kThree)^n$, with $n > 0$, can be approximated using the Schwinger parameterization \citep[\eg][]{Kenrick:2011:Basis}. 
Thus, many bispectra also have template-specific approaches to decomposing them. 
Instead of following different methods tailored to each given bispectrum, we follow the \textit{Planck} analyses in using mode decomposition methods on all templates, given we scan a wide range of model space. We now detail our chosen models below. 

We start with scale-dependent modifications to the local shape, given by \citep{Byrnes:2010:ScaleDepa}
\begin{mybox}    
\begin{align} \label{eqn:template:Local1}
    B^{\rm LSD, SF}(k_1,k_2,k_3) = \frac{(\kThree/k_\star)^{n_{\rm NG}}}{(\kOne\kTwo)^{3}} + 2~{\rm perm.}\,,
\end{align}
\end{mybox}
\noindent where $n_{\rm NG}$ is a power-law exponent and $k_\star = 1 \,\hinvmpc$ is a wavenumber (pivot) scale. This template corresponds to when the (scale-dependent) curvature perturbations are sourced by just one of the scalar fields. If multiple fields contribute to the perturbations, then we have
\begin{mybox}    
\begin{align} \label{eqn:template:Local2}
    B^{\rm LSD, MF}(k_1,k_2,k_3) = \bigg(\frac{\kOne\kTwo}{k_\star^2}\bigg)^{3 - n_{\rm NG}/2} + 2~{\rm perm.}
\end{align}
\end{mybox}
\citet{Byrnes:2010:ScaleDepa} generated these templates assuming $\log(k_{\rm max} / k_{\rm min})\, n_{\rm NG} \ll 1$. In our work, $\log(k_{\rm max} / k_{\rm min}) \approx 6$, so the templates are accurate up to $n_{\rm NG} \sim 0.1$. However, we treat this model as a phenomenological template for scale dependence and consider a wider range of $n_{\rm NG}$ values.

We then consider templates corresponding to resonant particle interactions. In this case, the inflaton potential has a localized feature that generates oscillations in the bispectrum \citep[see][for a review]{Achucarro2022InflationReview}. The models used in \citetalias{paper1} also exhibit oscillations, but are derived from explicit couplings of the inflaton to other massive fields \citep[\eg][]{Nima:2015:Colliders}, whereas the models we study here are described by phenomenological parameters alone. 
Sharp changes to the inflaton potential also induce oscillations at other orders, such as the power spectrum and trispectrum. 
In this work, we will first focus on oscillations only in the bispectrum, but then consider a class of models where the bispectrum \textit{and} power spectrum both contain oscillations. 
We ignore the primordial trispectra as its impact on late-time structure is subdominant to the power spectra and bispectra.\footnote{Note that there do exist some models where the primordial trispectrum can dominate over the bispectrum \citep{Philcox:2025:PaperI, Philcox:2025:PaperIII}. In general, the derivative of a signature in late-time structure, $X$, with the bispectrum amplitude $\fNL$ is $\mathcal{O}(10^3)$ higher than that with the trispectrum amplitude, $g_{\rm NL}$ \citep[\eg][see their ]{LoVerde:2011:trispec}.}

Oscillatory bispectra can be broadly classified into linear and logarithmic templates, given by \citep{Chen:2007:PNGs, Chen2010PNGReview}
\begin{mybox}    
\begin{align} \label{eqn:template:BkRes}
    B^{\rm LinRes}(k_1,k_2,k_3) & = \frac{1}{(\kOne\kTwo\kThree)^2} \cos[\omega (\kOne + \kTwo + \kThree) + \phi]\,, \\
    B^{\rm LogRes}(k_1,k_2,k_3) & = \frac{1}{(\kOne\kTwo\kThree)^2} \cos\left[\omega \log\left(\frac{\kOne + \kTwo + \kThree}{k_\star}\right) + \phi\right],
\end{align}
\end{mybox}
where $\omega$ is the frequency of the oscillations and $\phi = 0$ is the phase. We explore a fairly small range in $\omega \in [0.1, 2.5]$ relative to perturbation theory-based analyses of the CMB \citep[\eg][]{Beutler:2019:ResfNL}. Our choice is limited by the basis functions used in decomposing the templates. Oscillations of high order (large $w$) are not accurately reproduced by our existing basis functions.\footnote{In principle, this can be alleviated by using an adequately high order for the Legendre polynomials in Eq.~\eqref{eqn:modefunc}. However, doing so degrades the orthogonality of the basis set as discussed in Section 2.1 of \citetalias{paper1}. Higher values of $w$ could be explored by making improvements to the choice of basis functions used in the decomposition.}

As mentioned before, the oscillations in the bispectrum arise from sharp features in the inflaton potential. These features can also generate oscillations in the power spectrum. In Section \ref{sec:PspecBspec}, we will generate simulations that jointly include resonance signals in both the bispectrum and power spectrum. The signal in the latter is parameterized as
\begin{mybox}    
\begin{align} \label{eqn:template:PkRes}
    P^{\rm LinRes}(k) & = P(k) \, (1 + A_{\rm pk} \cos[\omega_{\rm pk} k + \phi_{\rm pk}])\,,\\[5pt]
    P^{\rm LogRes}(k) & = P(k) \, (1 + A_{\rm pk} \cos[\omega_{\rm pk} \log (k/k_{\star}) + \phi_{\rm pk}])\,.
\end{align}
\end{mybox}
where $P(k)$ is the original primordial power spectrum. In practice we set $\phi_{\rm pk} = 0$, and do not consider any phase offsets between the oscillations in the power spectrum and bispectrum. We use $k_{\star} = 1 \,\hinvmpc$ as with the bispectrum. Different physical processes can inject oscillations into the power spectrum, and various types have been studied with $N$-body simulations \citep[\eg][]{Schaeffer:2021:DarkOscillations, Stahl:2025:PkRes}. This work is the first to simultaneously simulate resonance-driven features consistently in the power spectrum and bispectrum.

However, the resonance bispectrum templates in Eq.~\eqref{eqn:template:BkRes} are an approximation \citep{Chen:2007:PNGs}. A more accurate analytic template, which also includes a damping envelope, is given by \citep{Adshead:2012:Kcos}
\begin{mybox}    
\begin{align} \label{eqn:template:Kres}
    B^{K^2 \cos}(k_1,k_2,k_3) & = \frac{1}{(\kOne\kTwo\kThree)^2}K^2 D(\alpha \omega K)\cos(\omega K)\,.
\end{align}
\end{mybox}
The model-dependent parameter $\alpha$ sets the maximum wavenumber cutoff, and $D(\alpha \omega K) = \alpha \omega/(K \sinh(\alpha \omega K))$ is a damping function. \citet{Adshead:2012:Kcos} also describe an analogous model, referred to as the ``$K\sin$'' model, which we do not consider in our work for brevity. The latter's features are broadly overlapping with the ``$K^2\cos$'' model we consider above. 

Bispectra can also be generated if the ground state of the inflaton is excited, which is also referred to as a non-Bunch-Davies (NBD) vacuum \citep[\eg][]{Chen:2007:single-field, Holman:2008:excited-initial, Meerburg:2009:initial-state}. The signal in these models generally peaks in the folded limit, $\kOne + \kTwo \approx \kThree$, though the exact behavior varies depeneding on the exact model under consideration. We only consider a subset of the NBD models studied in \citet{Planck:2014:PNGs, Planck:2016:PNGs, Planck2020PNGs} as the others were not accurately captured by our basis functions. Following the \textit{Planck} nomenclature, we explore the ``NBD mode 2'' model,
\begin{mybox}    
\begin{align} \label{eqn:template:NBD}
    B^{\rm NBD2}(k_1,k_2,k_3) & = \frac{1}{(\kOne\kTwo\kThree)^3}\bigg[(\kTwo\kThree)^2\times \frac{1 - \cos[(\kTwo + \kThree - \kOne)/k_c]}{\kTwo + \kThree - \kOne} + 2~{\rm perm.}\bigg]\,,
\end{align}
\end{mybox}
where the excitations are generated at conformal time, $\tau_c$ and set on a scale $k_c = -(\tau_c c_s)^{-1}$, with $c_s$ being the speed of sound. Note that this template does not diverge in the folded limit $\kTwo + \kThree - \kOne \rightarrow 0$ due to the $1 - \cos(\cdots)$ factor in the numerator; it is exactly zero at the folded configuration, but can exhibit a large (oscillatory) amplitude in its vicinity when $k_c$ is small.

Models with excited initial states are also often accompanied by resonant features. We use the template \citep{Chen:2010:NBD},
\begin{mybox}    
\begin{align} 
    B^{\rm NBD\sin}(k_1,k_2,k_3) & = \frac{1}{(\kOne\kTwo\kThree)^2}(e^{-\omega \tilde{k}_1} + e^{-\omega \tilde{k}_2} + e^{-\omega \tilde{k}_3}) \sin(\omega K + \phi)\,,
\end{align}
\end{mybox}
where $K = \kOne + \kTwo + \kThree$ and $\tilde{k}_1 = k_2 + k_3 - k_1$ with $\tilde{k}_2, \tilde{k}_3$ defined similarly. We set the phase $\phi = 0$ for all models.

Finally, we consider the template from Dirac-Born-Infeld (DBI) inflation \citep{Silverstein:2004:DBI, Alishahiha:2004:DBI},
\begin{mybox}    
\begin{align} \label{eqn:template:NBDsin}
    B^{\rm DBI}(k_1,k_2,k_3) = & \,\,\, \frac{1}{(\kOne\kTwo\kThree)^3} \frac{-3/7}{(\kOne + \kTwo + \kThree)^2}  \bigg[\big(\kOne^5 + 2~{\rm perm.}\big)  \nonumber\\  
    &+ \big(2\kOne^4\kTwo - 3\kOne^3\kTwo^2 + 5~{\rm perm.}\big) + \big(\kOne^3\kTwo\kThree - 4\kOne^2\kTwo^2\kThree + 5~{\rm perm.}\big)\bigg].
\end{align}
\end{mybox}
As discussed in the above works, this model arises from brane motion in warped extra dimensions, where a reduced sound speed enhances non-Gaussianity. 
The signal is highly correlated with the standard equilateral template.

Altogether, we consider thirty different templates in this work (varying the parameters of the templates above). Once we consider variations in the power spectrum as well, there are nearly fifty templates we study in total. 
As mentioned above, our choice of templates encompasses those studied by \textit{Planck} \citep{Planck:2014:PNGs, Planck:2016:PNGs, Planck2020PNGs}. 
Some of the templates (\eg the scale dependence, linear resonance, and NBD$\sin$) are manifestly separable and thus do not explicitly require a basis decomposition step. 
We still simulate these templates with our standard pipeline, both for consistency and because our pipeline will automatically suppress any relevant IR divergent terms in the bispectrum kernel.\footnote{\citet{Fondi:2025:Gengars} also provide an alternate prescription for suppressing IR divergences once provided with an arbitrary, separable kernel.} This broadly follows the approach of the \textit{Planck} analyses, where even simpler bispectra were analyzed with the mode decomposition approach.

\begin{figure}
    \centering
    \includegraphics[width=\columnwidth]{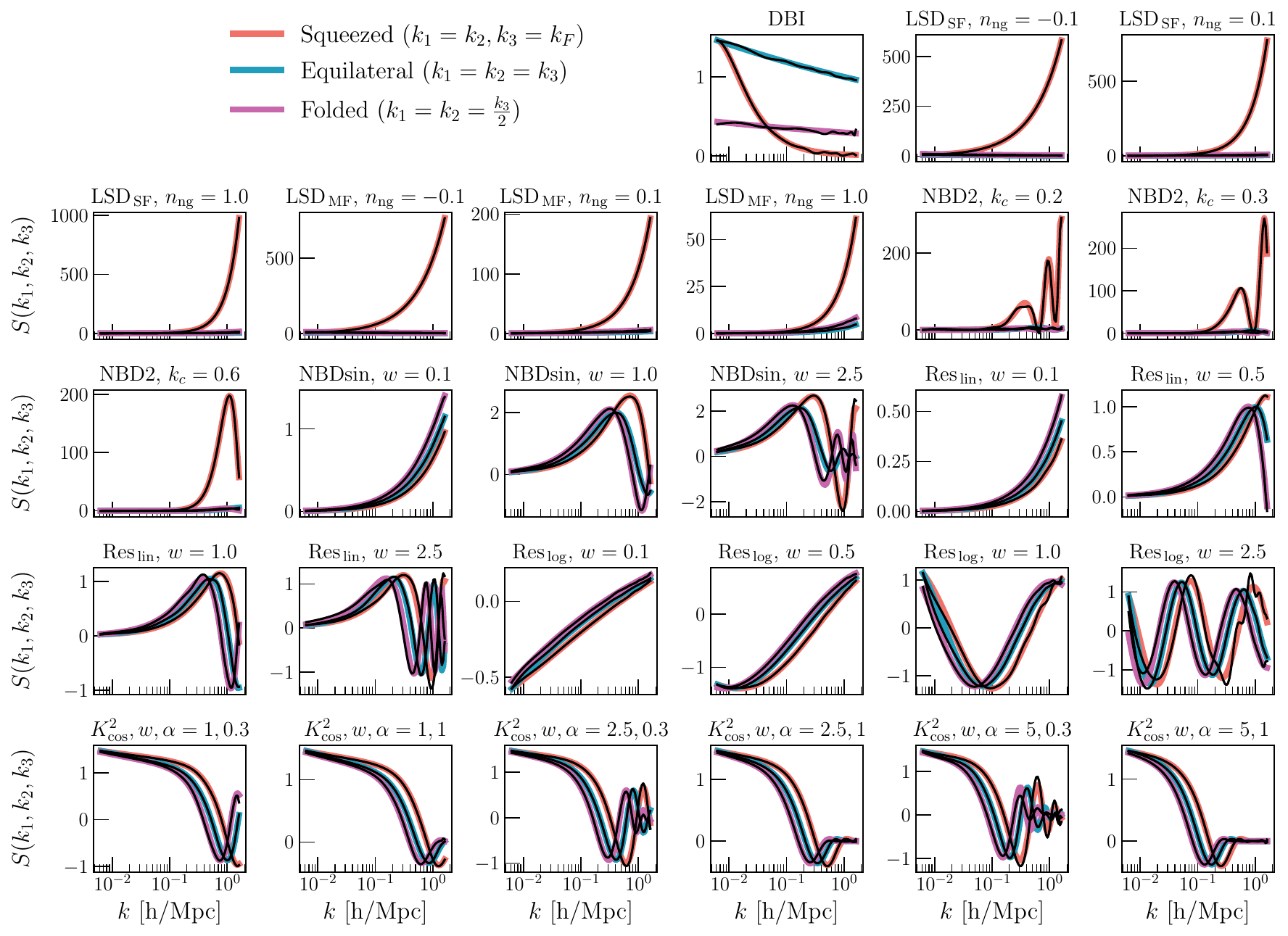}
    \caption{The different templates considered in this work, shown in three specific limits, alongside the approximated versions using our basis functions (black lines). The shape function is defined as $S(\ldots) = (\kOne\kTwo\kThree)^2\times B(\ldots) $. In all cases, the templates are adequately reproduced by our approximations; see Section \ref{sec:sims:Models} for details on the templates. Here, $k_{\rm F} = 0.006\, \hinvmpc$ is the fundamental frequency of the simulation volume.}
    \label{fig:Template}
\end{figure}

Figure \ref{fig:Template} shows the different templates that we consider in this work (colored lines), alongside the approximated versions using our basis functions (black lines). We show the squeezed, equilateral, and folded limits of the bispectrum. In all cases, the approximated template provides a close match to the true template, owing to the inherently oscillatory nature of the Legendre polynomials used in our basis. 
That said, our current basis functions do not reach sufficiently high order (for numerical reasons, see discussion above) to capture very high-frequency oscillations, which leaves room for improvement. 
All templates are simulated with $\fNL = \pm 100$. 
When considering models with oscillations in the bispectrum and power spectrum, we choose $\fNL = \pm 1000$. The latter choice further boosts the signal-to-noise of the derivative of a measured statistics with respect to $\fNL$ and helps pinpoint any subtle features from the interplay between power spectrum and bispectrum oscillations.

\section{Signature on Nonlinear Scales}\label{sec:results}

We first characterize the various bispectrum signals detailed above in how they impact nonlinear structure formation. In Section \ref{sec:PspecBspec} below, we do the same but for models where resonance signals are present in both bispectrum and power spectrum.

\subsection{Information in Weak Lensing Observations}\label{sec:results:lensing}

Measurements of weak lensing on smaller scales have been shown to have a strong sensitivity to $\fNL$ \citep{Shirasaki2012fNL, Anbajagane2023Inflation, paper1}. This is primarily because $\fNL$ alters the formation and evolution of massive structures, which dominate the signals on such scales. Conceptually, this signal can be viewed as originating from the impact of $\fNL$ on the halo mass function (HMF), which has been extensively documented by the community \citep{Dalal2008ScaleDependentBias, Shirasaki2012fNL, Marian2011fNL, Hilbert2012fNL, Jung2023fNLHMFQuijote}. However, a more precise statement is that the signal originates from changes to the statistics of \textit{peaks} in the density field, which have a clear dependence of $\fNL$ as detailed in \citet[][see their Appendix]{Dalal2008ScaleDependentBias}. See Section 5 of \citet{Anbajagane2023Inflation} for more detail on the origin of the $\fNL$ signal in weak lensing.

Following \citetalias{paper1}, we forecast constraints on the different PNGs using weak lensing measurements from the LSST Y10 dataset \citep{LSST2018SRD}. All choices in our analysis---including the forward model, covariance estimates, fisher matrix estimation, choice of summary statistic, etc.---follow that of \citetalias{paper1}, so we do not reproduce these descriptions here. The interested reader can find details of the forward model in Appendix D of \citetalias{paper1}, and of the summary statistics and Fisher forecast in Section 3.1 of the same work. 

In brief, we assume a survey footprint of 14,000 $\deg^2$ with a source galaxy number density of $n_{\rm gal} = 30\,\, {\rm arcmin}^{-2}$, following the LSST Year 10 (Y10) prescription. We use five tomographic bins as in \citet{Anbajagane2023Inflation}. Our summary statistics are the 2nd and 3rd moments of the lensing convergence field, $\kappa$. We measure all auto- and cross-correlations between the tomographic bins, and probe the scale dependence by smoothing the maps with 10 circular apertures between $3.2\arcmin < \theta < 200 \arcmin$. While our survey setup focuses on upcoming data (LSST Y10), these forecasts are still useful for existing datasets. A combination of lensing data from the Dark Energy Survey \citep{Gatti2021ShearCatalog, Secco2022Shear, Amon2022Shear} and the Dark Energy Camera All Data Everywhere (DECADE) project \citep{DECADE1, DECADE4, DECADE5} gives one access to 270 million galaxies spanning 13,000 $\deg^2$ of the sky, which is close to the same area we consider here. These existing galaxy samples probe lower redshifts \citep{Myles2021PhotoZ, DECADE2} than those anticipated from LSST Y10. However, our forecast here is still relevant as the signal is concentrated towards lower redshifts \citep[][see their Figure 6]{Anbajagane2023Inflation}.

The Fisher estimation requires two inputs: the covariance of our measured datavector and the derivative of the datavector with respect to the parameter of interest ($\fNL$). We estimate the former with 6000 independent realizations of the simulated datavector, leading to a Kaufman-Hartlap factor of $\gtrsim 0.95$ \citep{Kaufman1967, Hartlap2007} for our analysis, while the latter is estimated using 30 independent realizations. Derivatives contaminated by numerical noise can artificially break parameter degeneracies and thereby overestimate constraining power \citep{Coulton:2023:FisherTest}. In our case, we limit our key analyses to a single-parameter forecast on just $\fNL$.\footnote{Other approaches compress the datavector into a low-dimensional space where the impact of noise on the Fisher information is far less pronounced \citep{Coulton:2023:FisherTest}. However, defining the compression requires simulations as well, and given we only have 30 realizations for the $\fNL$ derivatives, we do not pursue this approach.} As a result, there is no such degeneracy breaking and the impact of numerical noise is subdued. We have explicitly confirmed our results are converged at the 2-3\% level if we estimate derivatives using half the available realizations.

\begin{figure}
    \centering
    \includegraphics[width=0.48\columnwidth]{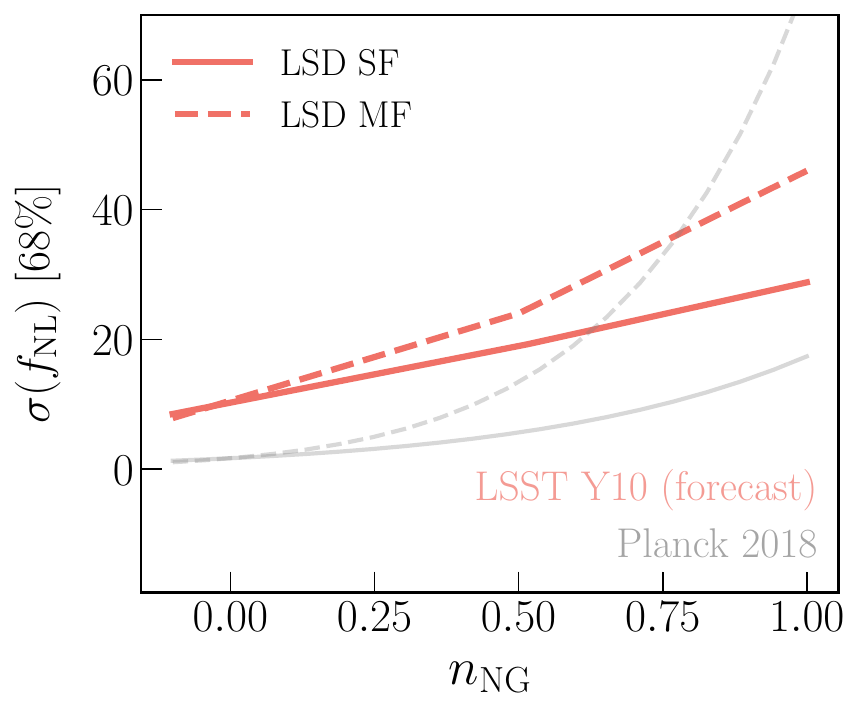}
    \includegraphics[width=0.49\columnwidth]{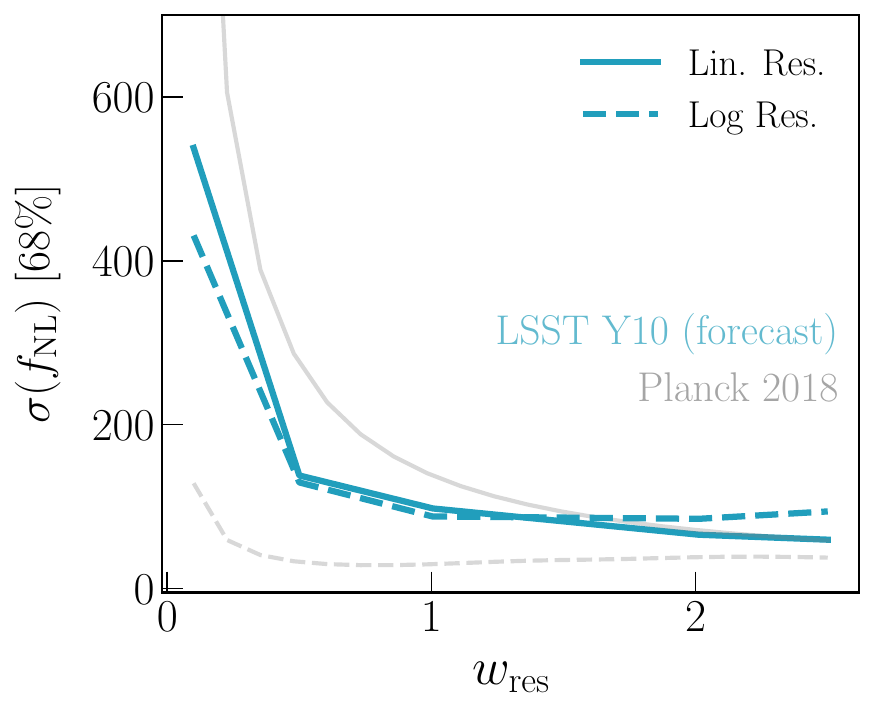}
    \includegraphics[width=0.49\columnwidth]{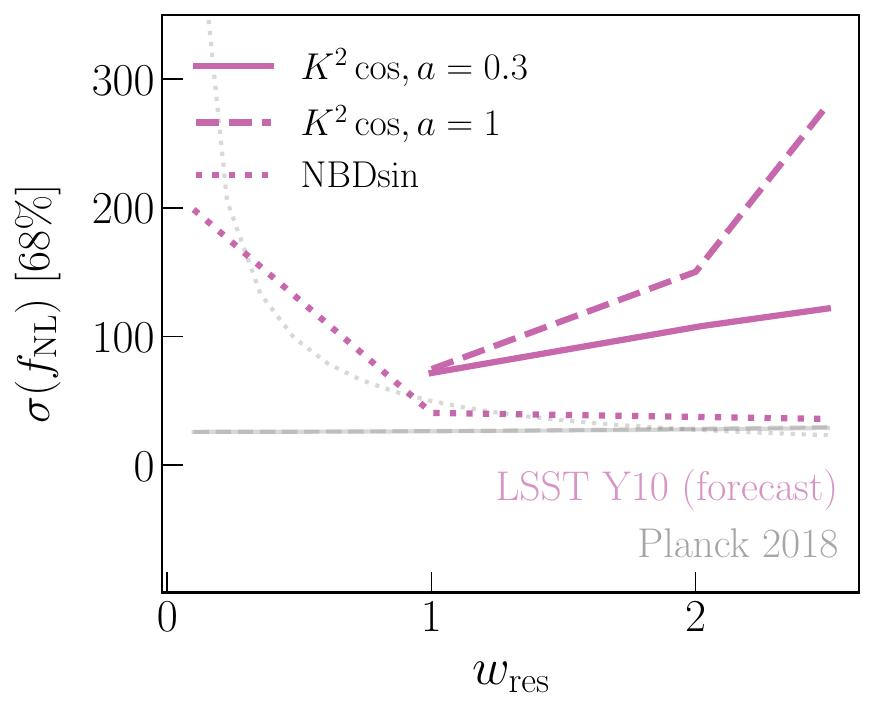}
    \includegraphics[width=0.48\columnwidth]{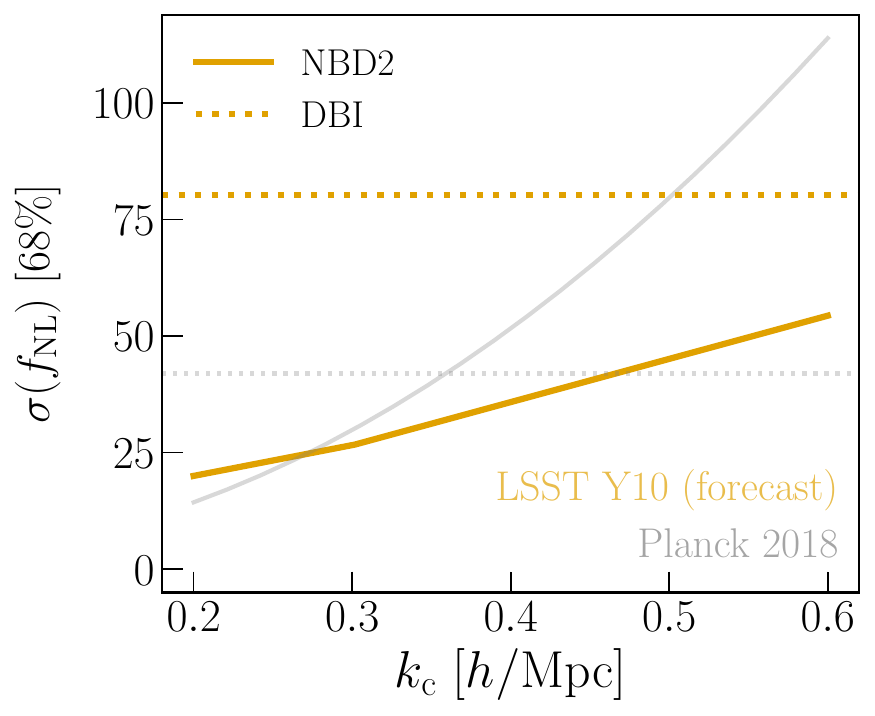}
    \caption{Constraints on the amplitude $\fNL$ for different models/templates and parameter spaces. Gray lines show our constraints from using the public CMB-BEST pipeline of \citet{Sohn:2024:Colliders}; see Appendix \ref{appx:Planck2018} for more details. The lensing constraints do not marginalize over additional parameters, where such marginalization is expected to degrade constraints by 20--30\%; see Section \ref{sec:results:lensing} for details. The forecasted lensing constraints are generally competitive with the data constraints of \textit{Planck} 2018, and can even surpass the latter for templates that have excess power on smaller scales. This highlights the synergy from using different probes with varying sensitivities.}
    \label{fig:Constraints}
\end{figure}

Figure \ref{fig:Constraints} presents the forecasts on various bispectrum models for LSST Y10 lensing data. Each model is characterized by the amplitude $\fNL$ and one or two additional parameters whose interpretations vary depending on the model. For each case, we generate simulations at selected values of the additional parameters, with the number of choices depending on the exact model, and compute the expected constraint on $\fNL$ for those fixed parameter values. 
In \citetalias{paper1}, we quantified the competitiveness of our forecasts by comparing them to the  \textit{Planck} 2018 constraints of \citet{Sohn:2024:Colliders}. 
While the models considered in this work similarly have constraints presented in \citet{Planck:2014:PNGs, Planck:2016:PNGs, Planck2020PNGs}, these constraints are generally quoted in terms of signal-to-noise and not $\sigma(\fNL)$. 
As such, no direct comparison with our forecasted uncertainties is available. 
We therefore perform our own analysis by utilizing the templates of Figure~\ref{fig:Template} on the temperature and polarization data from \textit{Planck} 2018.\footnote{Note that the combination of \textit{Planck} 2018 with the upcoming Simons Observatory data is expected to improve constraints by 20-50\% for the local, equilateral, and orthogonal templates \citep{SO:2019:Forecast}. We can then expect a similar level of improvement, relative to the \textit{Planck} 2018 constraints, for the templates studied here.} 
This is done using the publicly available CMB-BEST pipeline (and associated data products). We detail the analysis choices further in Appendix \ref{appx:Planck2018}.

The results of Figure \ref{fig:Constraints} show that, in general, the forecasted lensing constraints are competitive with the data constraints from \textit{Planck} 2018. In some cases, our forecasted results surpass the latter. 
This is expected as many of our templates have their signals peak on smaller scales. These scales are outside the range probed by \textit{Planck} ($2 \times 10^{-4} < k\, [\hinvmpc] < 0.2$) and are therefore not as easily constrained by it. Some templates, such as $K^2\cos$, peak towards large scales and explicitly damp the small-scale power; in these cases, the CMB naturally outperforms the lensing forecasts. 
Similarly, the CMB provides better constraints on our logarithmic (rather than linear) oscillatory template, since logarithmic oscillations span a wider range in $k$ and overlap more directly with the scales probed by the CMB. 

We note that our above statement, that structure-based constraints could outperform the CMB for signals peaking on small scales, is specific to the lensing analysis being discussed here. Conventional structure-based analyses of $\fNL$ \citep[\eg][]{Cabass2022MultifieldBOSS, Philcox2022BossPNG, Damico2022BossPNG} focus on the galaxy density field and use perturbative approaches to model both the galaxy bias and the spatial correlations in the matter density field. In this case there is a large uncertainty in (analytically) modeling the non-linear gravitational evolution of structure, and this significantly reduces the potential constraining power from using more non-linear scales. Such perturbative analyses are often limited to the same scales probed by CMB datasets, and therefore will not offer the same precision improvements discussed above. In the case of lensing, we only need to model the matter density field and therefore use simulation-based models, which can probe these smaller scales without incurring a significant penalty on constraining power. In principle, analyses of the galaxy density field can also avoid this penalty using simulations. However, the simulations needed for galaxy analyses are far more computationally expensive compared to their lensing counterparts \citep[][see their Section 5.1]{Anbajagane2023Inflation}.

In summary, Figure \ref{fig:Constraints} shows that upcoming lensing data can be competitive in constraining various primordial features, and could potentially be a leading probe for features whose signals peak on small scales. 
This highlights the synergy enabled by pursuing constraints on primordial physics from measurements of nonlinear structure formation.

As discussed in \citetalias{paper1}, our choice to perform single-parameter constraints---due to limitations in the simulation suite---means these results can be viewed as a best-case scenario when using the second and third moments. \citet{Anbajagane2023Inflation} show that marginalizing over additional parameters---such as $\sigma_8$, the root-mean squared amplitude of fluctuations at $z = 0$ on scales of $8 \mpc/h$, and $\Omega_m$, the fraction of energy density contributed by matter, as well as two nuisance parameters related to the intrinsic alignment of galaxies \citep{Troxel2015IAReview}---degrades constraints on $\fNL$ by 20\% to 30\% for the standard $\fNL$ templates (local, equilateral, and orthogonal). A similar degradation can be expected here as the templates we study are somewhat correlated with one or more of these standard templates. We note that existing works also limit their data constraints to a single-parameter fit when constraining $\fNL$ \citep[\eg][]{Planck2020PNGs, Sohn:2024:Colliders}.

\begin{figure}
    \centering
    \includegraphics[width=\columnwidth]{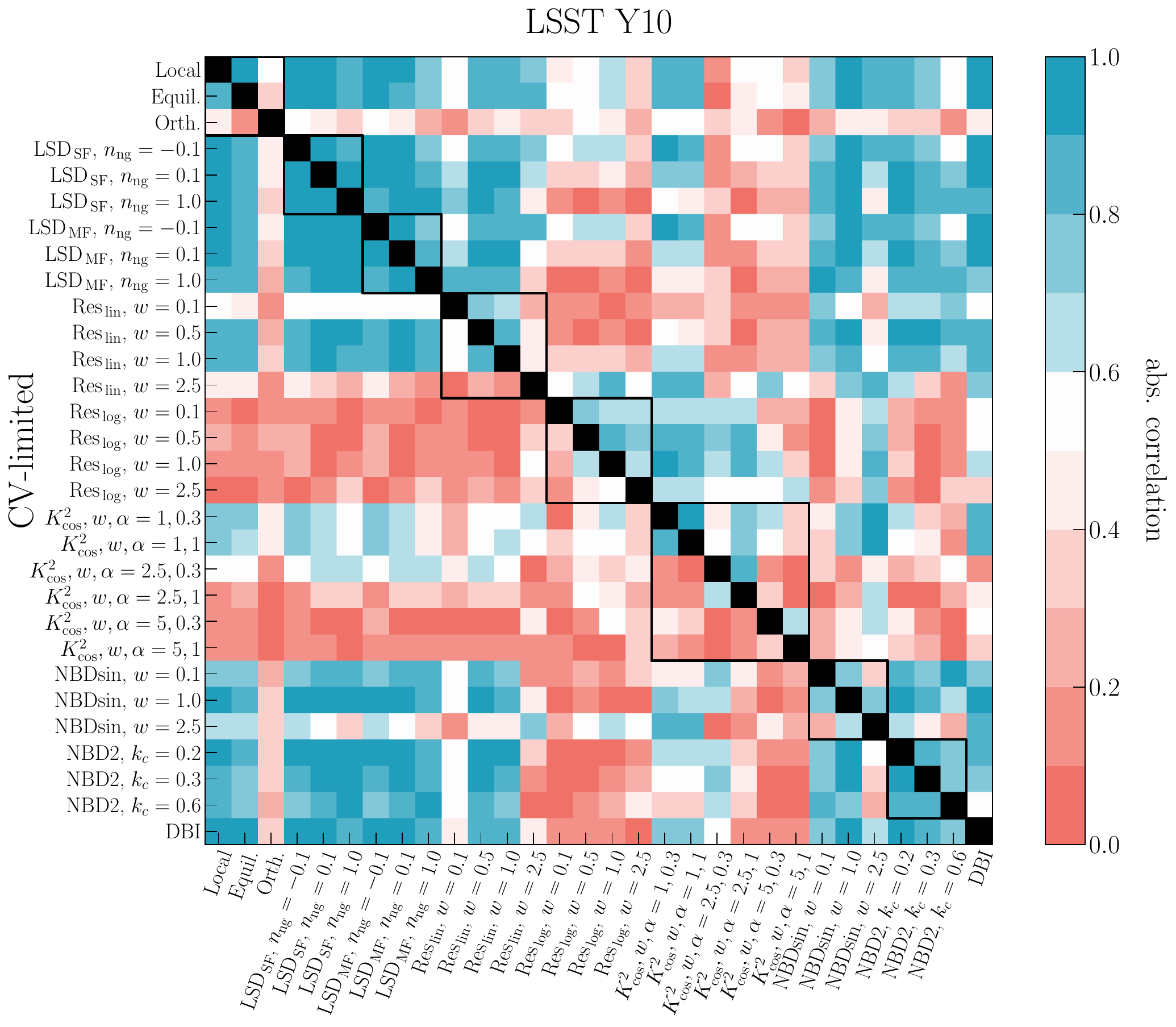}
    \caption{The correlation in the joint Fisher posterior for pairs of models. Values close to 1 and 0 indicate strong degeneracy and orthogonality, respectively, between the late-time predictions of two models. The upper triangle uses the parameter covariance matrix for LSST Y10, and the lower one the same for a cosmic variance (CV) limited survey. Redder colors indicate more uncorrelated signals. The correlations in an LSST Y10 measurement are similar to those from a CV-limited survey. The diagonal (which has a trivial correlation of 1) is shaded black for clearer visuals. The black outlines denote different classes of templates.}
    \label{fig:Corr}
\end{figure}

Figure \ref{fig:Corr} presents the cross-correlation of the late-time signatures of different templates, as seen in the second and third moments of the lensing convergence field. We compute this by estimating the Fisher information of a two-model analysis,\footnote{As discussed in \citetalias{paper1}, deviating from a single-parameter analysis can cause our estimates to be impacted by noise in the derivatives \citep{Coulton:2023:FisherTest}. We have confirmed that our estimates of the off-diagonal covariance changes by at most $\Delta{\rm corr} = 0.3$ if we use one-third as many realizations. This does not change our qualitative discussions about the correlation between two templates. We do not make any quantitative statements.} and evaluating the parameter covariance matrix. We normalize the off-diagonal component to obtain the correlation, and take the absolute value to isolate just the amplitude (and not sign) of the correlation. 
Values close to 1 indicate strong correlations between the late-time predictions of two templates, and those close to 0 indicate strong orthogonality between the same. 
The upper triangle shows results for datavectors in an LSST Y10 survey, while the lower triangle is the same for datavectors in a cosmic variance-limited survey, i.e., one with no shape noise in the lensing estimates. The black boxes approximately delineate different classes of the templates discussed in Section~\ref{sec:sims:Models}.

The models with resonances have a clear de-correlation compared to the scale-dependent models and the NBD models. The class of models corresponding to resonant particle production is broadly orthogonal to the other models considered in this work, and is therefore a useful addition. The NBD models, while exhibiting oscillatory behaviors, are still fairly correlated with the local and equilateral ones. The resonance models also exhibit weak correlations between templates that use different choices of $w$. This indicates it may be possible to jointly constrain $w$ and $\fNL$ for these models. We still follow the approach of \citetalias{paper1} in constraining $\fNL$ given a set of fixed values for the other physical parameters of the model.

\subsection{Matter Power Spectrum and Bispectrum} \label{sec:results:matter}

Following \citetalias{paper1}, we showcase the features in the late-time, nonlinear matter power spectrum and the bispectrum under a certain PNG template. We compute the bispectrum using the estimator of \citet{Scoccimarro:2015:Bispectrum} as detailed further in Appendix B1 of \citetalias{paper1}.

\begin{figure}
    \centering
    \includegraphics[width=\columnwidth]{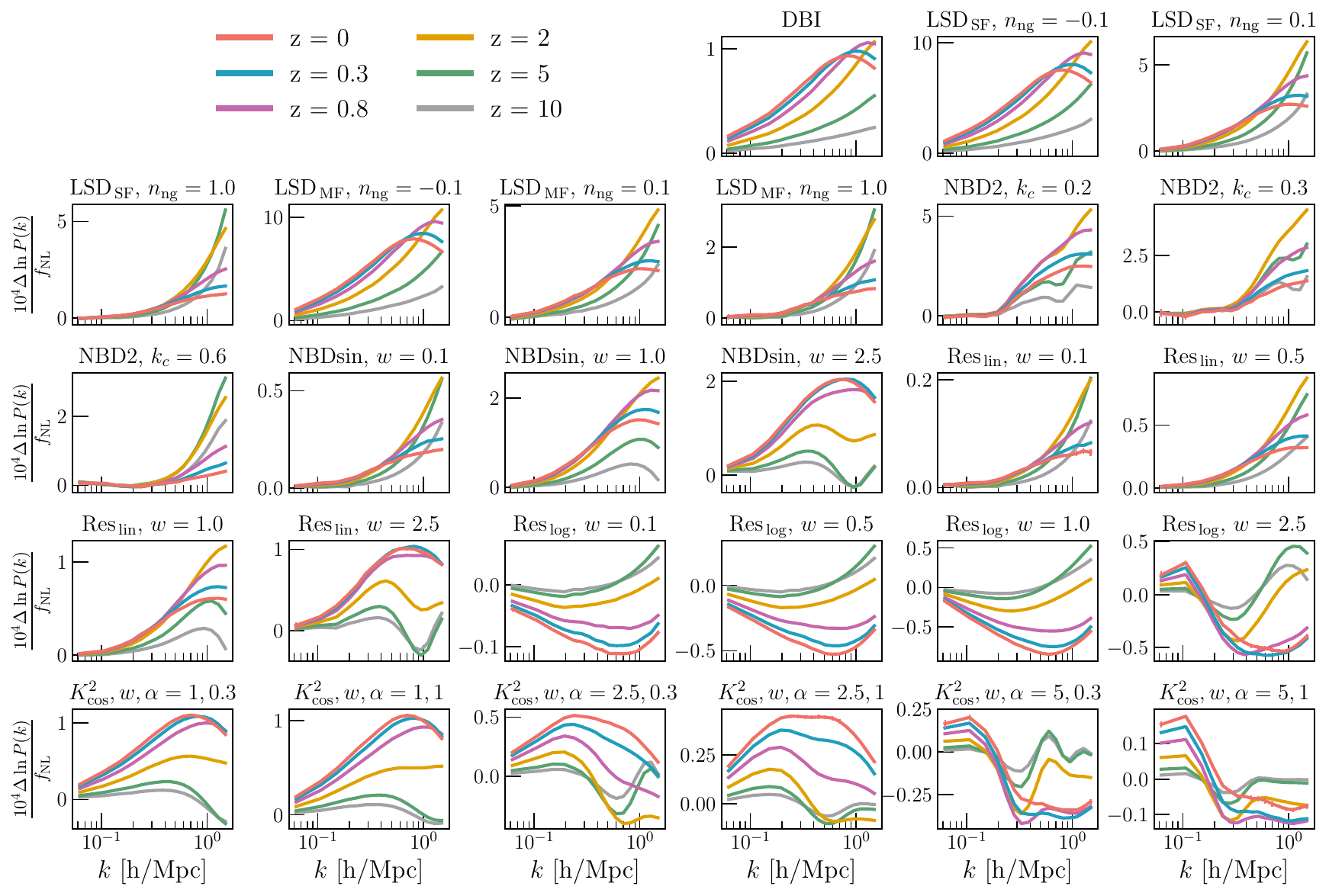}
    \caption{The derivative of the power spectrum with input $\fNL$ value, presented for different templates and at different redshifts. For values of $\fNL = 100$, there is a 1\% to 2\% change in the power spectrum across all scales. Oscillatory features are more prevalent at high redshift, and are washed away at lower redshifts as nonlinear structure evolves over time. However, the amplitude of the derivatives still increases towards lower redshifts with the prolonged impacts of nonlinear structure formation. The error bars show the uncertainty on the mean derivative, obtained from bootstrapping over ten independent realizations. For visibility, we only show uncertainties at $z = 0$. Though, in most cases the error bars are hidden behind the lines.}
    \label{fig:results:mPk}
\end{figure}

Figure \ref{fig:results:mPk} presents the fractional change in the power spectrum for unit $\fNL$. For characteristic values of $\fNL = 100$, the change in the power spectrum is $\approx 1\%$. Some models, such as the scale-dependent variants of the local PNGs, exhibit significantly larger changes. The NBD2 models show a clear signal beyond $k \gtrsim k_c$, where the template shows nontrivial behavior (Figure \ref{fig:Template}), and a nearly mean-zero derivative below that scale. The resonance models, particularly the linear resonances, show clear oscillatory features in the power spectrum. Note that in this analysis the primordial power spectrum is a simple power law (with no oscillations). So any oscillations observed in the late-time power spectrum arise from the primordial bispectrum's influence on the evolution of the power spectrum. 
In general, the oscillatory behavior is still prevalent at $z = 2$---which is a redshift regime that is accessible with current and future surveys \citep{Spergel2013, Euclid, LSST2018SRD}---and diminishes significantly at lower redshifts. The latter occurs as nonlinear structure formation (the formation of filaments, the collapse of halos, etc.) modifies the primordial features. This coupling of primordial features and nonlinear structure has also been explored more in \citet{Goldstein:2025:CosmoColl}.

\begin{figure}
    \centering
    \includegraphics[width=1\columnwidth]{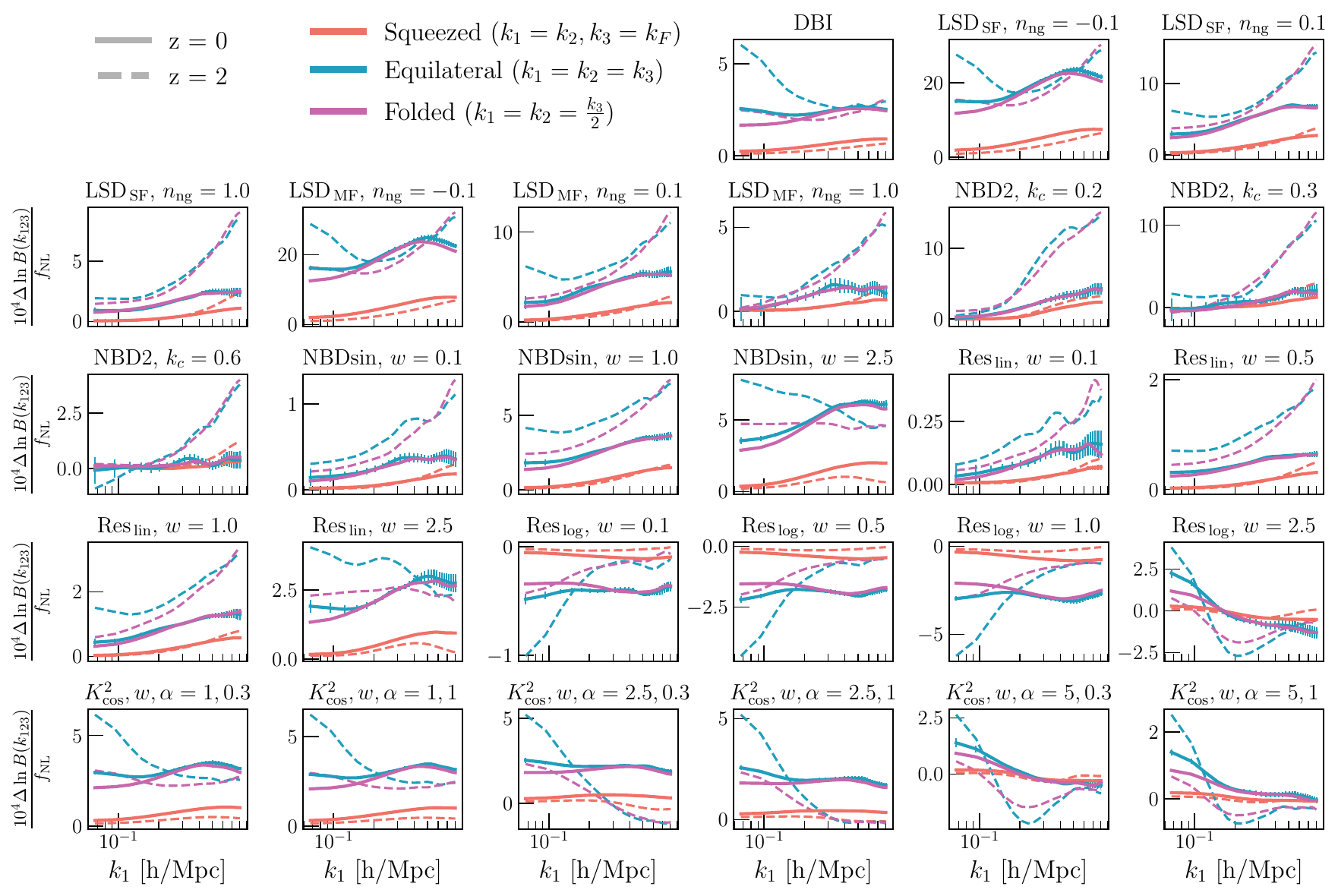}
    \caption{Similar to Figure \ref{fig:results:mPk} but for the matter density bispectrum. For visibility reasons, we only show uncertainties for the equilateral and squeezed limits at $z = 0$. The measurements are smoothed with a narrow Gaussian kernel for visualization.}
    \label{fig:results:mBk}
\end{figure}

Figure \ref{fig:results:mBk} then presents the fractional change in the bispectrum for unit $\fNL$. As discussed in \citetalias{paper1}, these differences are largest in the equilateral and folded limits as the squeezed limit of the late-time bispectrum is dominated by the contributions from gravitational nonlinearities \citep{Takahashi:2020:BiHaloFit}, so the primordial component has a much smaller relative contribution. The oscillations of the linear resonance models are less pronounced (compared to the power spectrum signal in Figure \ref{fig:results:mPk}) but those of the log resonances and the $K^2\cos$ models can be observed better. The scale-dependent models show a clear excess in signal on small scales as we increase $n_{\rm NG}$. More positive values of this power-law index increases (decreases) the amplitude of PNGs on small (large) scales. However, this excess is significantly reduced as we approach $z = 0$. Once again, the results are consistent with nonlinear structure modifying the primordial signal.

\subsection{Halo Abundance and Bias} \label{sec:results:halos}

We now check the impact of PNGs on statistics of massive halos. While PNGs impact halos over a range of mass scales, the resolution of our simulation limits us to studying only objects with $\Mtwohc > 10^{13.5} \msun/h$ as these are resolved by at least 30 particles. This resolution is adequate for our target of measuring halo counts and halo clustering. The analysis in this section follows that of \citet{Anbajagane2023Inflation} and \citetalias{paper1} who identified the formation of massive halos as the primary origin for why the weak lensing field is sensitive to $\fNL$, in agreement with earlier work on the PNG information content in halo statistics \citep{Dalal2008ScaleDependentBias, Shirasaki2012fNL, Marian2011fNL, Hilbert2012fNL, Jung2023fNLHMFQuijote}.

\begin{figure}
    \centering
    \includegraphics[width=\columnwidth]{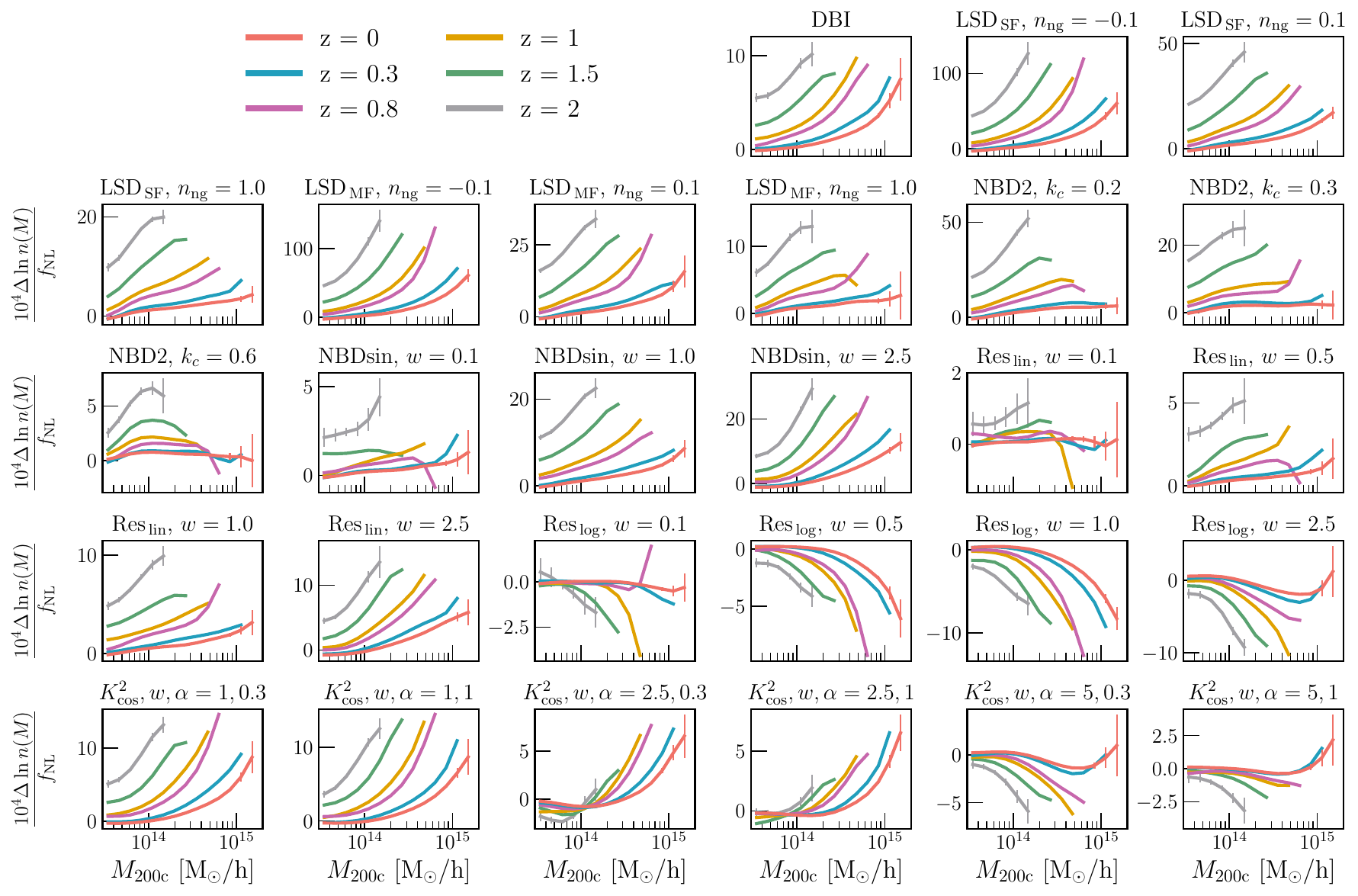}
    \caption{Similar to Figure \ref{fig:results:mPk} but for the halo mass function (HMF). A value of $\fNL = 100$ results in an order 10\% to 25\% change in the halo mass function in many models. At fixed mass, the derivatives scale with redshift as the peak height (or ``rarity'') increases. For visibility, we smooth the measurements with a narrow Gaussian kernel and also show the uncertainties only at $z = 0$ and $z = 2$. A number of oscillatory models exhibit non-monotonic changes, where halos of a particular mass are more sensitive to a given PNG template compared to those of smaller/larger masses.}
    \label{fig:results:HMF}
\end{figure}

Figure \ref{fig:results:HMF} presents the change in the halo mass function (HMF) for a unit change in $\fNL$. As shown in \citet{Anbajagane2023Inflation}, the sensitivity of the HMF to PNGs is an order of magnitude larger than the other effects discussed above. Following the arguments of prior work \citep[\eg][\citetalias{paper1}]{Dalal2008ScaleDependentBias, Shirasaki2012fNL, Anbajagane2023Inflation}, we note that PNGs change the shape of density field's 1-point distribution. This change preferentially up/down-weights the tails of the distribution, and the abundance of massive halos is sensitive to these tails \citep{Press:1974:HMF}. As a result, the HMF changes significantly as we vary the amplitude of PNGs. Following the same argument, the most massive halos (whose formation depends on the more extreme/low-probability regime of the distribution) are more significantly impacted by a PNG template relative to their less-massive counterparts.

Our results show that the formation of halos is, as expected, sensitive to power on large scales. The scale-dependent models with $n_{\rm NG} < 0$ show a significant boost to the HMF at all masses. Increasing the index to $n_{\rm NG} = 1$ reduces the impact by an order of magnitude as now the template's power peaks on smaller scales.\footnote{While massive halos have a dominant impact on the small scale clustering of the late time density field, their formation is sensitive to a significantly larger patch of the initial density field. For a halo of $\Mtwohc = 10^{15} \msun/h$, the (comoving) Lagrangian diameter is $30 \mpc/h$, which is around $k = 2\pi/30 \approx  0.2 \hinvmpc$. Given we choose a pivot of $k_\star = 1 \hinvmpc$ (see Eqs.~\ref{eqn:template:Local1} and \ref{eqn:template:Local2}) variations in $n_{\rm NG}$ will modulate the power on these scales.} A number of templates show non-monotonic features---they preferentially boost the HMF for a narrow range of halo masses and only weakly alter the HMF for more/less massive halos. Examples are the NBD2, log resonance, and $K^2\cos$ models. The latter shows a particularly interesting morphology, where the HMF at $z < 0.5$ is preferentially suppressed for halos of $\Mtwohc = 5 \times 10^{14} \msun/h$. The exact mass scale can be altered by varying the phase of the oscillation (which is set to $\phi = 0$ in the default template of \citet{Adshead:2012:Kcos}). We highlight this in the context of cluster cosmology \citep[see][for a review]{Allen:2011:Clusters}, where current results indicate that higher mass samples prefer a slightly lower value of $\sigma_8$ (and therefore lower abundance of such halos) relative to lower mass cluster samples \citep[][see their Figure 7]{DES:2025:6x2ptY3}.

\begin{figure}
    \centering
    \includegraphics[width=\columnwidth]{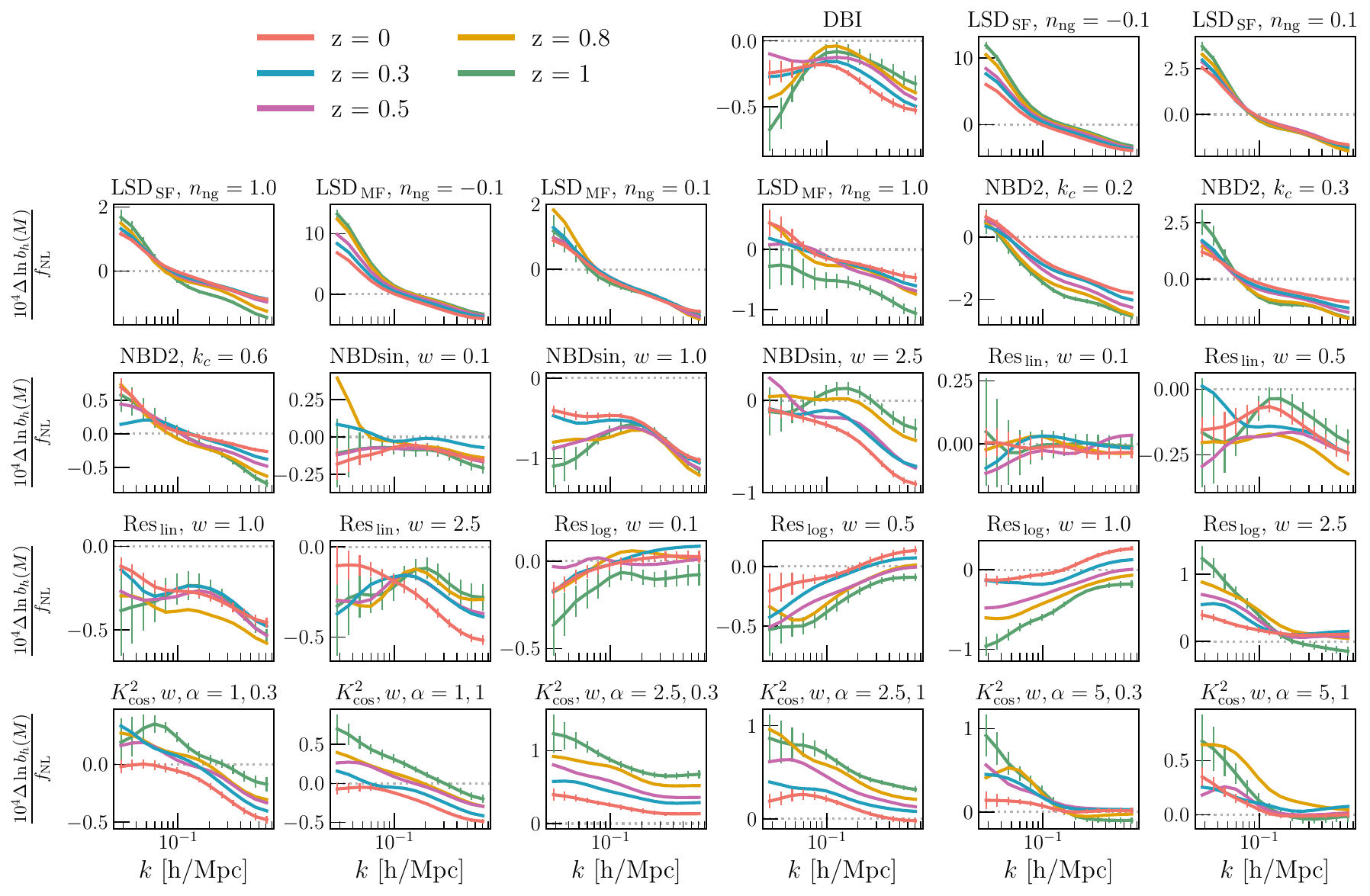}
    \caption{Similar to Figure \ref{fig:results:mPk} but for the halo bias, estimated as $P_{\rm hm}/P_{\rm mm}$. We estimate this for a sample with $M_{\rm 200c} > 10^{14} M_\odot/h$. We show the (bootstrap-derived) errors at $z = 0$ and $z = 1$ to enhance visibility. The measurements are also smoothed with a narrow Gaussian kernel for visualization.}
    \label{fig:results:halobias}
\end{figure}

Finally, Figure \ref{fig:results:halobias} presents the change in halo bias for unit change in $\fNL$. We estimate this using all halos above $\Mtwohc > 10^{14} \msun/h$, and through the ratio $b(k) = P_{\rm hm}(k)/P_{\rm mm}$. Here $P_{\rm hm}(k)$ and $P_{\rm mm}$ are the halo-matter cross-spectrum and matter-matter auto-spectrum, respectively. We have confirmed that our estimates are not sensitive to shot noise, by recalculating the bias using a randomly subsampled half of the halo catalog and confirming the resulting estimate is consistent with our fiducial results. The amplitude of the measured halo bias generally increases with redshift, as the bias increases with halo peak height \citep[\eg][]{Tinker2010Bias} and at fixed mass threshold the peak height increases with redshift. The scale-dependent variations to the local template result in halos exhibiting the well-known scale-dependent bias \citep{Dalal2008ScaleDependentBias}, albeit with a slightly different scaling since $n_{\rm NG} \neq 0$. For the PNGs sourced by resonant particle production, we find the halo bias exhibits non-monotonic behaviors. The clearest features occur for larger values of $w$ and are diminished as we move to $z = 0$. As expected, the templates that generate a non-monotonic difference in the HMF also generate non-monotonic differences in the halo bias. Focusing on all templates that are highly correlated with local PNGs, we note that the derivative of halo bias with $\fNL$ exhibits a clear zero-crossing. Such a feature was noted for the standard local type in \citet[][see their Figure 8]{Anbajagane2023Inflation}. Other resonance templates, particularly for higher values of $w$, also exhibit such zero-crossings.

\subsection{Interplay Between the Primordial Bispectrum and Power Spectrum}\label{sec:PspecBspec}

The resonance models considered in the above sections can generate oscillations not only in the bispectrum, but in the power spectrum as well. For the remainder of this section, we use the notation $w_{\rm bk}$ and $w_{\rm pk}$ to distinguish the bispectrum and power spectrum oscillation frequencies.

We consider a specific scenario, where the bispectrum has a fixed oscillation, $w_{\rm bk} = 2$, and we scan a range of power spectrum oscillation frequencies, $\wpk \in [2, 8, 64]$. The former is slightly smaller than the highest frequency whose corresponding template we can reliably decompose using our basis functions (Figure \ref{fig:Template}). We incorporate oscillations into the power spectrum model using Eq.~\eqref{eqn:template:PkRes}. The primordial power spectrum is renormalized so that $\sigma_8 = 0.834$, consistent with the fiducial value of all simulations in the \textsc{Ulagam} suite \citep[][\citetalias{paper1}]{Anbajagane2023Inflation}. We perform a two-parameter Fisher forecast of the power spectrum and bispectrum amplitudes, $\Apk$ and $\fNL$ respectively. Our analysis considers the linear resonance and log resonance models from Section \ref{sec:sims:Models}. In both cases, the power spectrum oscillations are also linear/logarithmic.

\begin{figure}
    \centering
    \includegraphics[width=\columnwidth]{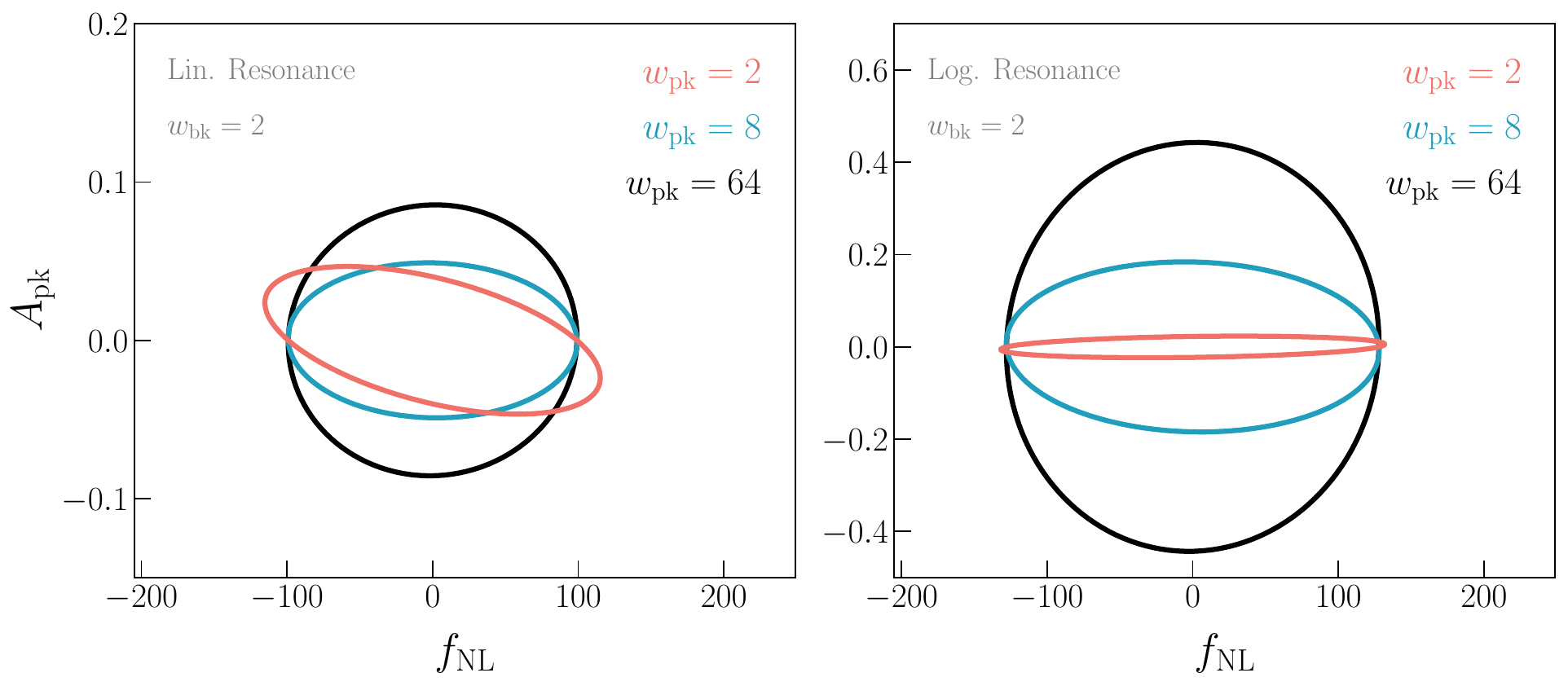}
    \caption{Joint constraints ($1\sigma$ contours) on the amplitude of the bispectrum ($\fNL$) and power spectrum ($\Apk$) for resonance models with oscillations in both spectra. We consider linear (left) and logarithmic (right) oscillations; see Section \ref{sec:sims:Models}. In all cases, the bispectrum has a fixed oscillatory frequency of $w_{\rm bk} = 2$, and we vary the frequency of power spectrum oscillations. In the linear case, there is a mild degeneracy between parameters that reduces as we increase the frequency of the power spectrum oscillations. There is no such degeneracy in the logarithmic resonance model.}
    \label{fig:Constraints_PkBk}
\end{figure}

Figure \ref{fig:Constraints_PkBk} shows the $1\sigma$ forecasts on the amplitudes. There is only a mild degeneracy between parameters, and present only in the linear resonance model when using lower values of $\wpk$. The log resonance model finds no degeneracy between the parameters. This indicates that analyses of bispectrum resonant signals can be done independently without needing to include power spectrum oscillations. However, we caveat that we have only explored a small region of a potentially wide parameter space, and are considering a simple Fisher information metric that does not capture nonlinear parameter correlations. Nonetheless, the results of Figure \ref{fig:Constraints_PkBk} are promising for the goal of independently constraining oscillations in both spectra. We have confirmed our constraints here are numerically converged, as the quoted parameter posterior widths change by only $\approx 5\%$ if we use half of our simulations to estimate the Fisher information.

\begin{figure}
    \centering
    \includegraphics[width=\columnwidth]{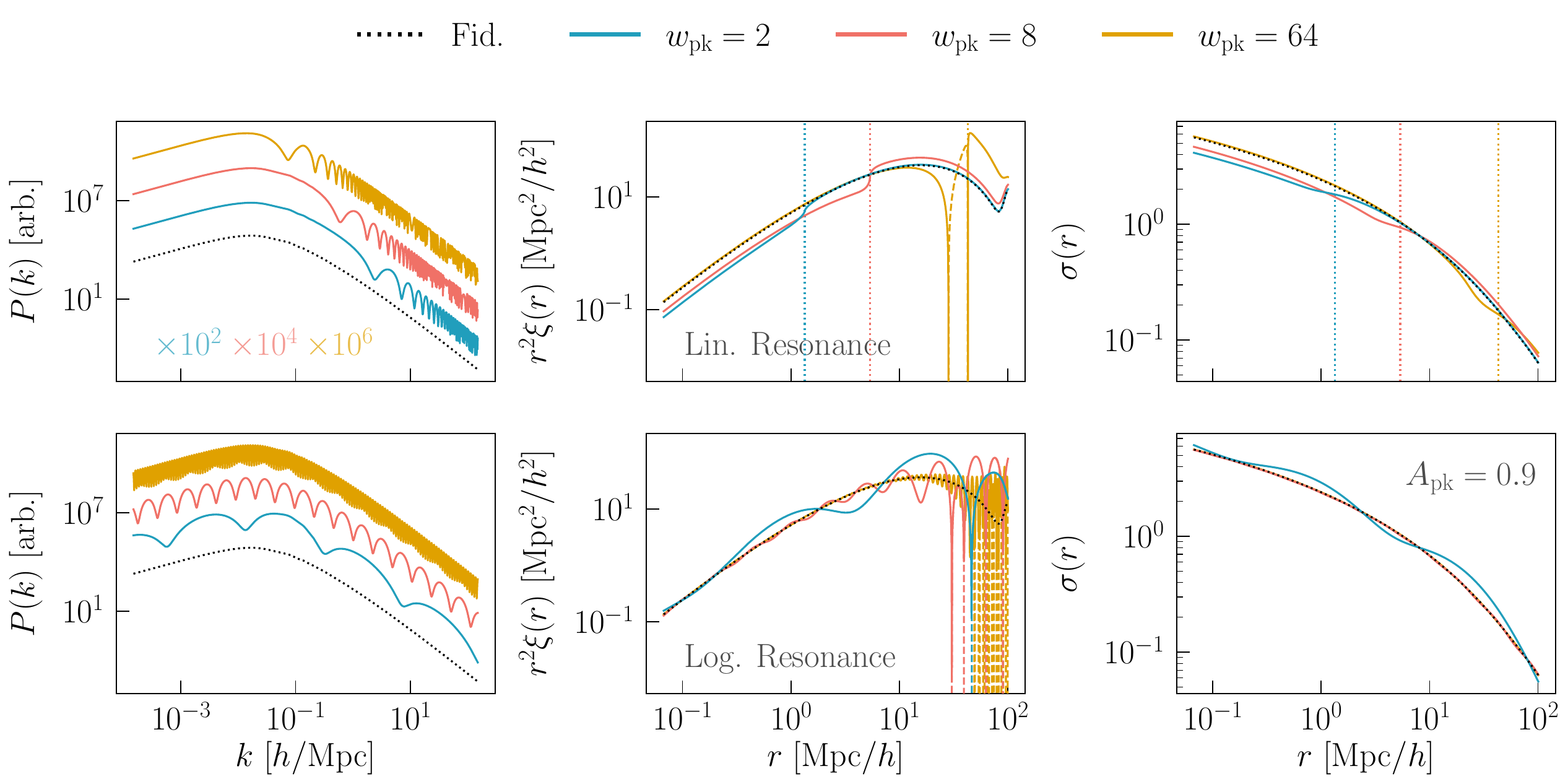}
    \caption{The linear and logarithmic resonance signals in the $z = 0$ linear matter power spectrum (left), the corresponding two-point matter correlation function (middle), and the standard deviation of density fluctuations within circular apertures (right), which is the generalized analog of the quantity $\sigma_8$ but now for length-scale $r$. We multiply the power spectra by arbitrary values to separate them for visual purposes. All results use $\Apk = 0.9$. Dashed lines denote where the function is negative. The linear model causes a clear deficit (and/or excess) at a specific physical scale, $r = w$, denoted in the dotted lines. The log resonance model, on the other hand, produces the same correlation function but now with oscillations atop it. The overall effect is indiscernible for $\wpk \gtrsim 8$. This results in the reduced sensitivity of the lensing data to this log resonance model.}
    \label{fig:results:xi_resPk}
\end{figure}

The constraints on $\Apk$ are also interesting on their own, as a single-parameter analysis. Forecasts of linear resonances generally consider $\wpk \gtrsim 10$ \citep[][see their Figure 10]{Beutler:2019:Pkres} due to modeling challenges. For example, the forecasted uncertainty from galaxy clustering probes increases by an order of magnitude below $\lesssim 30 \mpc$, due to the challenge of modeling the galaxy field on such scales using perturbation theory. According to \citet{Beutler:2019:Pkres}, Stage IV spectroscopic surveys will achieve $\sigma(\Apk) \sim 0.3$. 
When considering only lensing data, we find $w = 8$ could be constrained at $\sigma(\Apk) = 0.03$, which is an order of magnitude better. However, the situation is inverted for larger frequencies---our constraints for $w = 64$ are $\sigma(\Apk) = 0.05$, whereas the aforementioned surveys achieve $\sigma(\Apk) = 0.001$ in this regime. Similar to Section \ref{sec:results:lensing} above, we find lensing is most powerful when constraining features that originate on small scales.

Appendix \ref{appx:PsecBspecExtra} provides a more detailed description of how power spectrum oscillations alter nonlinear structure formation. For intuition on the nature of our constraints in Figure \ref{fig:Constraints_PkBk}, we now provide a simple example using linear theory. Figure \ref{fig:results:xi_resPk} shows the resonance-modified signal in the linear matter power spectrum (left), the corresponding real-space correlation function (middle), and the standard deviation of the matter density fluctuations (right). The latter is connected to the formation of massive halos and is useful for understanding how these oscillations will impact halo abundances \citep{Press:1974:HMF}. In all cases, we rescale the power spectrum (after the feature has been added) such that it produces $\sigma_8 = 0.834$. All calculations are performed using the \textsc{Core Cosmology Library} \citep{Chisari2019CCL}.

In the linear resonance model, the frequency $\wpk$ can be interpreted as a real-space length scale (in units of $\mpc/h$). A corresponding feature is present at those scales (denoted by dotted colored lines) in the two-point correlation function. There is a transition in power above/below a certain length scale, consistent with the discussions of other studies on oscillatory power spectra and structure formation \citep[\eg][]{Stahl:2025:PkRes}. This, in turn, is a signal that lensing data is sensitive to and can therefore place constraints on. In the case of the log resonance models, the resulting $\xi(r)$ quantity matches the fiducial result but with oscillatory residuals atop it. For lower values of $\wpk$, the results still mimic some scale dependence (which allows lensing to place tighter constraints as seen in Figure \ref{fig:Constraints_PkBk}). However, for larger values of $\wpk$ the correlation more closely mimics the fiducial result, and this causes a vanishing sensitivity of the signal in lensing data. This is also seen more clearly in $\sigma(r)$ which shows no sensitivity to the log resonances above a certain frequency.

In summary, Figure \ref{fig:results:xi_resPk} identifies why lensing data is more sensitive to $\wpk \lesssim 30$ and why it is more sensitive to linear resonance models than logarithmic resonance ones. Finally, we note these results use linear theory (without any nonlinear extensions or simulations) and are only meant to illustrate the main features of the underlying signal. The more rigorous analysis of these signals, folding in the full nonlinear behavior via $N$-body simulations, is performed in Appendix \ref{appx:PsecBspecExtra}. Previous works have also studied different models of power spectrum oscillations using $N$-body simulations \citep[\eg][]{Schaeffer:2021:DarkOscillations, Stahl:2025:PkRes}.

\section{Conclusions}\label{sec:conclusions}

In this work, we employ our novel method for generating initial conditions with arbitrary bispectra \citepalias{paper1} and generate the first suite of simulations that propagate the impact of thirty different primordial bispectra---considering a variety of signatures such as resonances and excited initial states---into the deeply nonlinear regime of structure formation. Following \citetalias{paper1}, we also provide extensive validation of our initial conditions method (Appendix \ref{appx:Validation}) and showcase its robustness from numerical artifacts. In addition, we also study a class of resonance models where the features are imprinted consistently in both the power spectrum and bispectrum.

Our main results are summarized as follows:

\begin{itemize}
    \item Lensing can provide constraints on $\fNL$ that are competitive to \textit{Planck} 2018, with the potential to surpass it for PNG models where the signal peaks on small-scales near/beyond the edge of \textit{Planck}'s sensitivity (Figure \ref{fig:Constraints}). All CMB constraints in this work are derived using the CMB-BEST framework (Appendix \ref{appx:Planck2018}).
    
    \item Consistent with other work, the primary signature of PNGs in the nonlinear density field is through changes to the abundance of massive halos (Figure \ref{fig:results:HMF}, and Figure \ref{fig:results:halobias}). Models with oscillatory bispectra imprint non-monotonic features in these relations.
    
    \item The late-time matter power spectrum and bispectrum are sensitive to changes in $\fNL$ (Figure \ref{fig:results:mPk} and Figure \ref{fig:results:mBk}). There are a number of non-monotonic features in the power spectra, arising purely from the presence of such features in the bispectrum. The morphology of these features is generally suppressed towards late times ($z \rightarrow 0$) due to nonlinear structure formation. However, the overall signal in the density field continues to grow with redshift; it is only the non-monotonic morphology that is suppressed with time.

    \item In models with resonances in both power spectrum and bispectrum, we find the constraints on their respective amplitudes are independent (non-degenerate) with one another (Figure \ref{fig:Constraints_PkBk}). Lensing is a powerful probe of small-scale linear oscillations in the power spectrum, which is a regime not well-constrained with existing data. We detail the full sensitivity of nonlinear structure to power spectrum oscillations in Appendix \ref{appx:PsecBspecExtra}.
\end{itemize}

$N$-body simulations have been an ever-present tool in the analysis of density fluctuations in the late Universe. Their uses in our fiducial cosmology analyses range from model testing and validation \citep[\eg][]{DeRose:2019:Buzzard, To:2024:Cardinal}, to semi-analytic extensions into more nonlinear scales \citep[\eg][]{Takahashi2012Halofit, Takahashi:2020:BiHaloFit}, and to full forward-modeled simulation-based inference \citep[\eg][]{Zurcher2022WLPeaks, Fluri2022wCDMKIDS, Gatti:2024:WPH, Prat:2025:Homology}. Different aspects of these techniques have been used extensively to model probes like weak lensing, cluster abundance, galaxy clustering etc. \citep[\eg][]{Krause2021Methods, DECADE3, To:2025:ModellingY6, Gomes:2025:Map3}. However, such techniques have generally not been viable for studies of inflation---and particularly for studies beyond the simpler inflationary models---due to the lack of simulation suites that include this signal. We have improved on this in \citetalias{paper1}, by providing a framework for generating simulations with arbitrary bispectra in their initial conditions. In that work, we utilized this method to show the utility of late-time, nonlinear probes in studying the signals from particle interactions in the primordial Universe; the so-called ``cosmological collider physics'' models. In this work, we have shown the same techniques enable us to study a wider range of models, and highlight how nonlinear structure can be more than competitive in constraining the model parameter spaces.

The ushering in of simulation-based techniques---which have been extensively used and tested in analyses of $\Lambda$CDM and other extended models---will enable studies of inflation to enter a multi-probe landscape, where constraints are informed by (and cross-checked between) multiple classes of observations, each with its own advantages and caveats. Many cosmological analyses have benefited tremendously from having access to multiple such comparison points, and the same will be true for inflation as well. This series of works builds on existing efforts in the community to provide the necessary tools to enable such a future.

\section*{Acknowledgements}

DA thanks many people---Austin Joyce, Josh Frieman, Lucas Secco, Olivier Dore, Ben Wandelt, Lam Hui, Sam Goldstein, Oliver Philcox, Chihway Chang, Bhuvnesh Jain, Wayne Hu, Katrinn Heitmann, Salman Habib, Will Coulton, Paco Villaescusa-Navarro, Chun-Hao To, Nick Kokron, Moritz Munchmeyer, and Mat Madhavacheril---for enlightening conversations across the two years leading up to the publication of this work. DA is supported by the National Science Foundation Graduate Research Fellowship under Grant No. DGE 1746045. HL is supported in part by the DOE award DE-SC0013528.

All analysis in this work was enabled greatly by the following software: \textsc{Pandas} \citep{Mckinney2011pandas}, \textsc{NumPy} \citep{vanderWalt2011Numpy}, \textsc{SciPy} \citep{Virtanen2020Scipy}, and \textsc{Matplotlib} \citep{Hunter2007Matplotlib}. We have also used
the Astrophysics Data Service (\href{https://ui.adsabs.harvard.edu/}{ADS}) and \href{https://arxiv.org/}{\texttt{arXiv}} preprint repository extensively during this project and the writing of the paper.

\section*{Data Availability}

Our pipeline for generating initial conditions---including our technique for approximating bispectra using separable terms---is available at \url{https://github.com/DhayaaAnbajagane/Aarambam}. The simulations are publicly released as part of the \textsc{Ulagam} suite. More details can be found at \url{https://ulagam-simulations.readthedocs.io}. The released products include the 2D density shells, 3D density fields, and halo catalogs for all snapshots. Please contact DA if you are interested in other products from the simulations.

\bibliographystyle{mnras}
\bibliography{References}

%%%%%%%%%%%%%%%%%%%%%%%%%%%%%%%%%%%%%%%%%%%%%%%%%%

%%%%%%%%%%%%%%%%% APPENDICES %%%%%%%%%%%%%%%%%%%%%

\appendix

\newpage
\section{Additional Results from Power Spectrum and Bispectrum Resonances}\label{appx:PsecBspecExtra}

In this appendix, we extract the features in the matter field and halo field using simulations that include oscillations in the primordial power spectrum.

\subsection{Matter Power Spectrum} \label{sec:results:matter}

\begin{figure}
    \centering
    \includegraphics[width=\columnwidth]{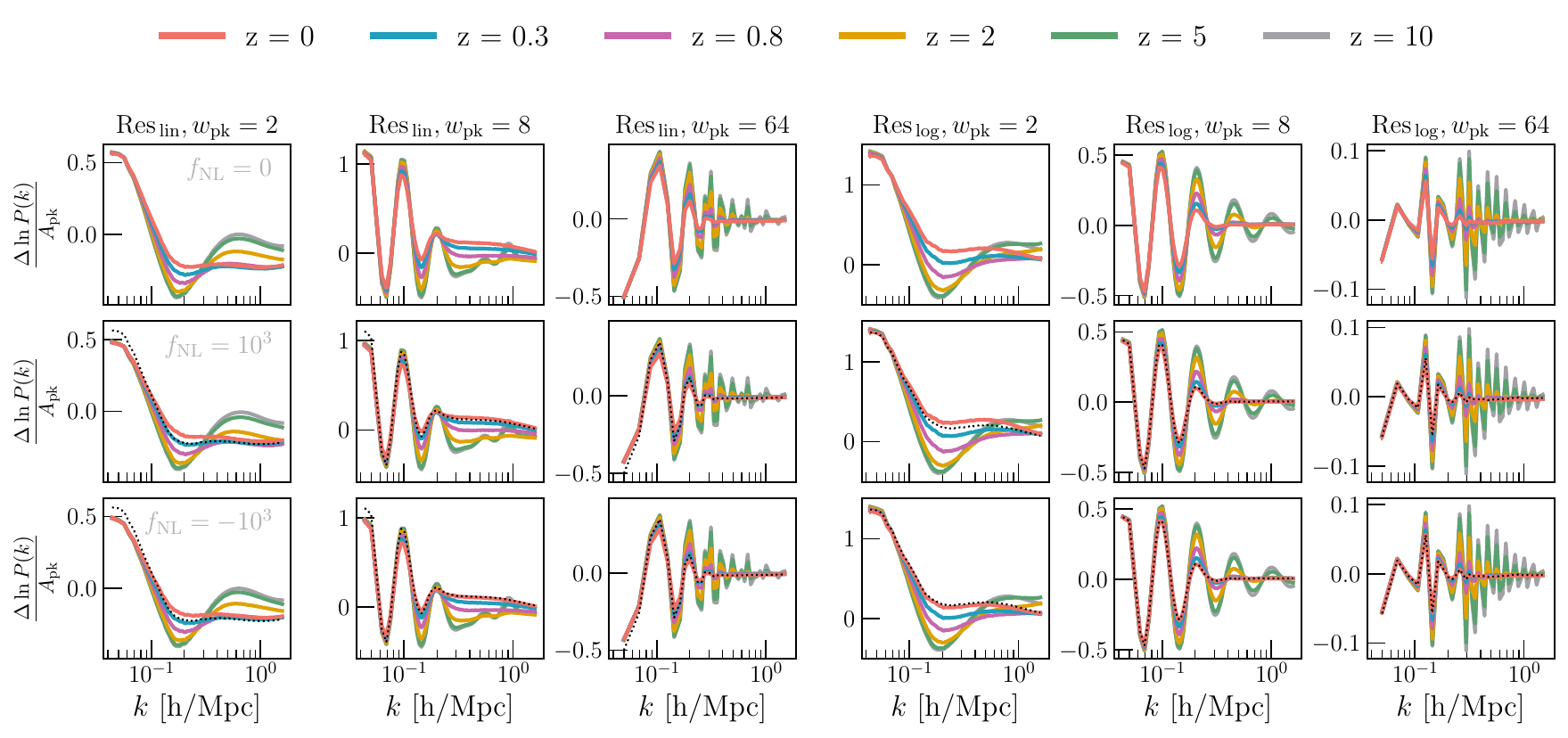}
    \caption{The derivative of the power spectrum with $\Apk$, the amplitude of power spectrum oscillations, for linear/log resonances, and at different redshifts. We show results up to $z = 10$ to provide a baseline for the features present prior to nonlinear structure formation. All oscillatory features are suppressed in structure at $z = 0$, but only on scales of $k > 0.2 \hinvmpc$. Larger-scale features are still preserved at late times. The $z = 0$ derivative shows associated error bars (on the mean, estimated from 10 independent realizations), but they are not visible on this scale as the signal dominates the uncertainties.}
    \label{fig:results:mPkApk}
\end{figure}

Figure \ref{fig:results:mPkApk} shows the change in the matter power spectrum when we vary the amplitude of the oscillatory features in the primordial power spectrum; see Eq.~\eqref{eqn:template:PkRes}. As expected there are $\mathcal{O}(50\%)$ changes with an oscillatory pattern consistent with the frequency of the primordial signal. We find clear signs that features on $k < 0.2 \hinvmpc$ are mostly unaffected by nonlinear structure formation, while those above this scale are rapidly suppressed as $z \rightarrow 0$, due to the impact of nonlinear structure. Note that Figure \ref{fig:results:mPkApk} does not rescale the derivative by any factors---the parameter $\Apk$ has an $\mathcal{O}(1)$ impact on the late-time matter power spectrum. In the case of the log resonance model, the effect is significantly diminished as we increase the resonance frequency. Figure \ref{fig:Constraints_PkBk} already shows that the constraints on $\Apk$ widen by a factor of 20 as we go from $w = 2 \rightarrow 64$, and the results on the matter power spectrum corroborate that. Figure \ref{fig:results:xi_resPk} above presents a simple exercise that explains the origin of this behavior.

The different rows in Figure \ref{fig:results:mPkApk} show the derivatives in simulations whose initial conditions correspond to different values of  $\fNL$. The black dotted line in the lower two rows are the $z = 0$ and $\fNL = 0$ result from the top row. We see there is a visible difference (with changes of $\Delta \fNL = 1000$), but it is mild and consistent with the decoupled nature of $\fNL$ and $\Apk$ signals indicated in Figure \ref{fig:Constraints_PkBk}. That is, a nonzero primordial bispectrum does not noticeably affect the derivative of the power spectrum with respect to $\Apk$. 

The bispectrum shows largely the same features as the power spectrum---oscillations that increase with $\wpk$ and that are suppressed towards lower redshifts, with the log resonance models showing poorer sensitivity like in other results presented thus far. For brevity, we do not showcase the results here.

\subsection{Halo Abundance and Bias} \label{sec:results:halos}

\begin{figure}
    \centering
    \includegraphics[width=\columnwidth]{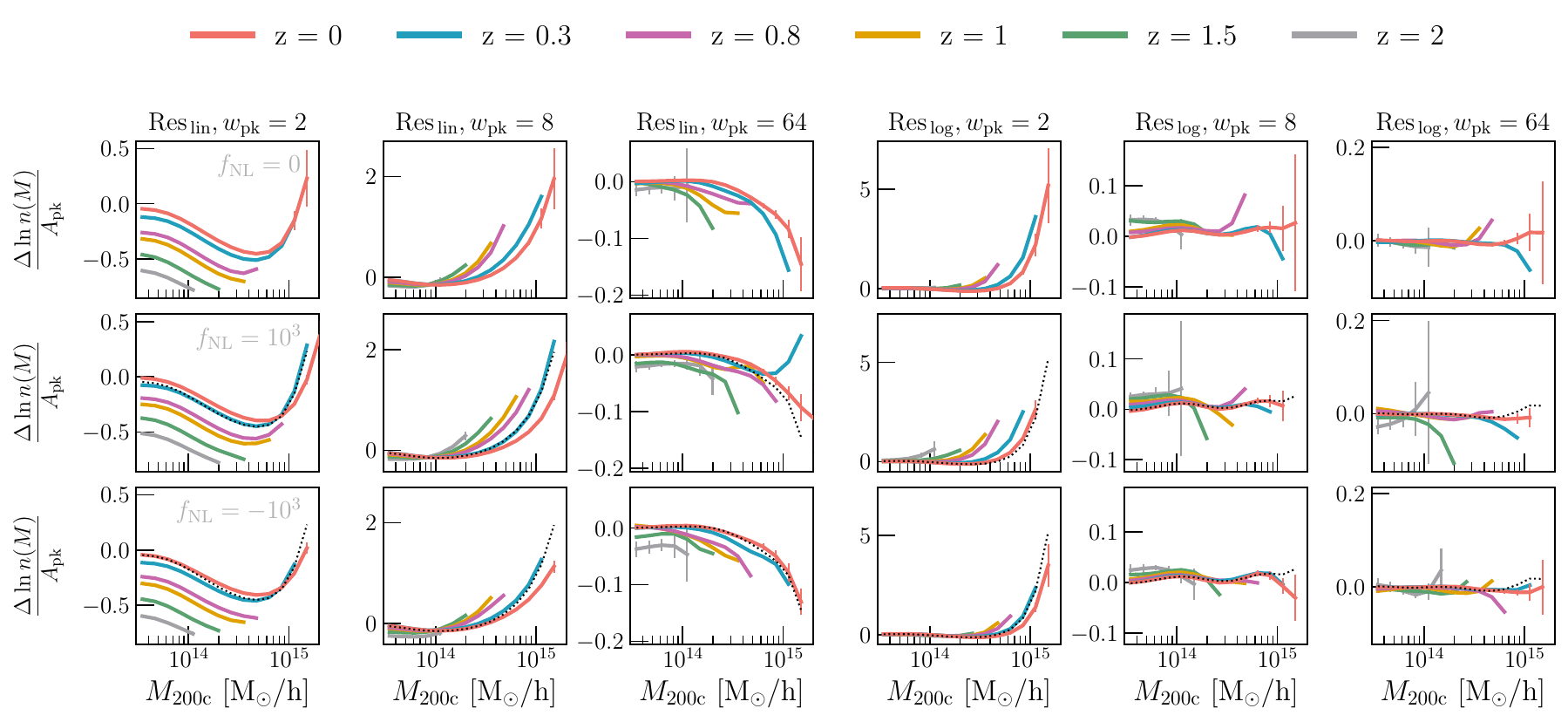}
    \caption{Similar to Figure \ref{fig:results:mPkApk} but for the halo mass function (HMF). There are mild, non-monotonic features in the HMF for the linear resonance model. These are suppressed for larger values of $\wpk$. The log resonance model shows little impact on the HMF.}
    \label{fig:results:HMFApk}
\end{figure}

The HMF shows clear, non-monotonic features as a function of halo mass. Focusing on the linear resonance models, we see that increasing $\wpk$ causes the derivative to be suppressed for less massive halos. In the logarithmic models, the derivatives are highly suppressed and essentially zero. Following \citet{Press:1974:HMF}, the HMF depends on the integral of the power spectra, i.e., the $\sigma(r)$ quantity presented in the right panels of Figure \ref{fig:results:xi_resPk}. Rapid oscillations in $P(k)$ get averaged out in this integral and result in little-to-no sensitivity of the HMF to the resonance signals.

\begin{figure}
    \centering
    \includegraphics[width=\columnwidth]{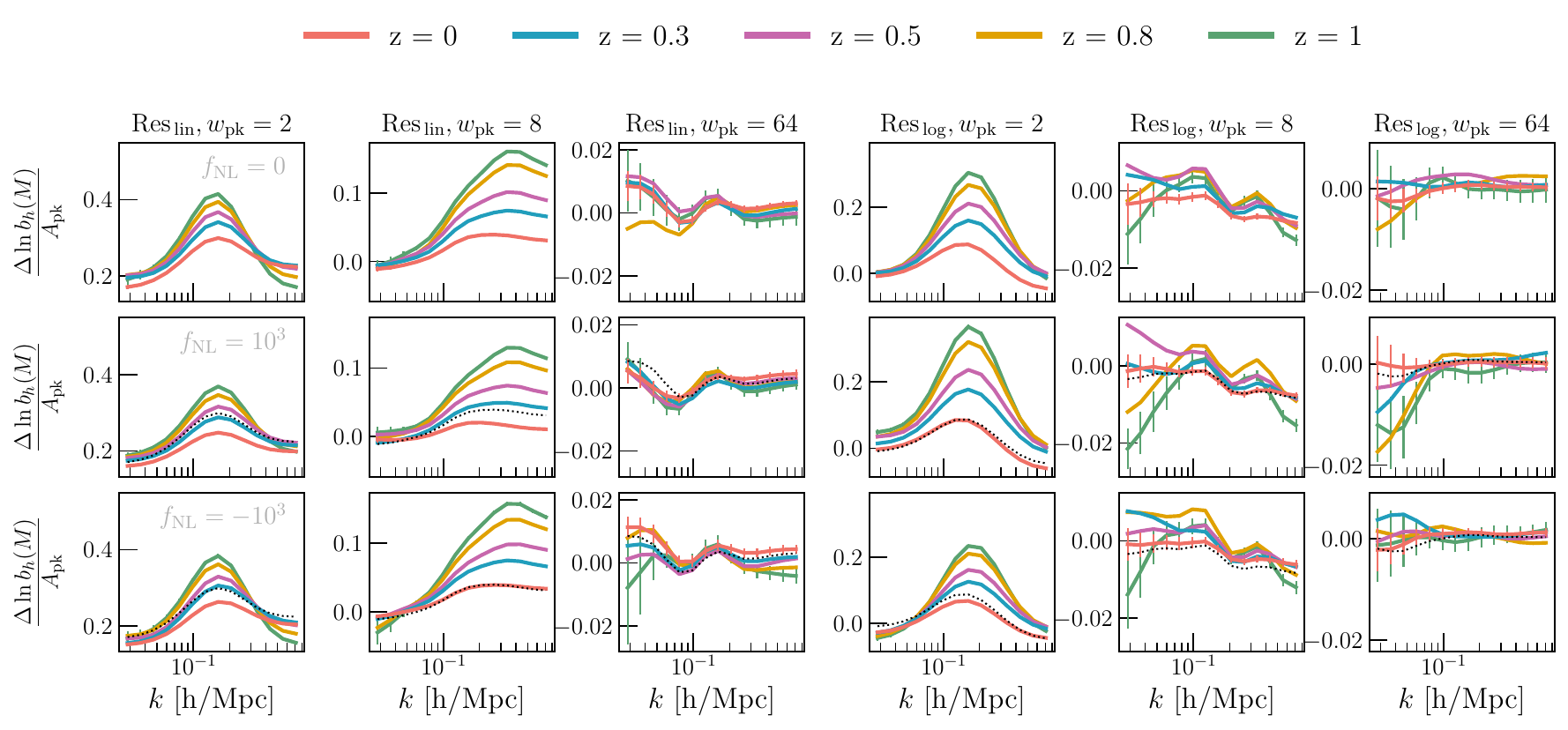}
    \caption{Similar to Figure \ref{fig:results:mPkApk} but for the halo bias, estimated as $P_{\rm hm}/P_{\rm mm}$, the ratio of the halo-matter cross-spectra and the matter auto-spectra.}
    \label{fig:results:halobiasApk}
\end{figure}

A similar argument can be made for the halo bias presented in Figure \ref{fig:results:halobiasApk}. The data show vanishing sensitivity to the linear resonance signal as we increase $\wpk$, since the resonance feature now gets pushed to larger scales, $r$. The log resonance model, when considering higher values of $\wpk$, shows little to no impact on the halo bias. This is in contrast to the bispectrum resonance model, where the bias does show a clear dependence on this signal (Figure \ref{fig:results:halobias}). In general, we find the halo bias behaves in a similar manner to the HMF---there are some mild non-monotic features, but these are suppressed as we increase $\wpk$.

\section{Constraints from \textit{Planck} 2018}\label{appx:Planck2018}

In Section \ref{sec:results}, we forecast constraints on inflationary models using LSST Y10 lensing data. To gauge the competitiveness of  these constraints, we require a comparison point. While all models we study have an analogous analysis performed in the \textit{Planck} data \citep{Planck:2014:PNGs, Planck:2016:PNGs, Planck2020PNGs}, the results are often quoted in units of signal-to-noise. Many constraints do not provide an estimate of the uncertainty, $\sigma(\fNL)$.

We circumvent this limitation by performing our own CMB-based analysis. For this, we leverage the public CMB-BEST pipeline\footnote{\url{https://github.com/Wuhyun/CMB-BEST}} provided by \citet{Sohn:2024:Colliders}. Section 3.2 of this work details an efficient and compact estimator for $\fNL$ given the temperature and polarization maps from \textit{Planck} 2018. For our analysis, we simply pass the templates from Figure \ref{fig:Template} to CMB-BEST. These templates are then projected into the different basis functions, and the coefficients of the terms are used alongside pre-computed data products in CMB-BEST to estimate $\fNL$. Similar to the flagship \textit{Planck} analyses, \citet{Sohn:2024:Colliders} utilize a simulation-based method for estimating uncertainties on $\fNL$. We have confirmed that our pipeline reproduces the constraints presented in \citet{Sohn:2024:Colliders}, and that our results for standard templates (local, equilateral, orthogonal) are consistent with the fiducial results of \citet{Planck2020PNGs}.

We highlight two specific details on this analysis. The first is that the public CMB-BEST pipeline utilizes the bispectrum template within the range $2 \times 10^{-4} < k [\!\hinvmpc] < 0.2$.\footnote{The coefficients $\beta_n$ and $\gamma_{nn^\prime}$ in Eq.~(3.13) of \citet{Sohn:2023:CMBbest} are precomputed assuming a given set of basis functions and a given $k$-range. We utilize the provided coefficients and do not recompute them ourselves for different choices of the basis function and of the scale range.} This is different from the range used in our work above, $6 \times 10^{-3} < k [\!\hinvmpc] < 3$, which is set by the size and resolution of the simulation. The mismatch is expected as it emerges from the nature of the different probes and highlights their complementarity. The second detail is that this analysis uses the original CMB-BEST pipeline with the original basis function choices. That is, we do not use the modified basis function set presented in \citetalias{paper1} and discussed above in our main analysis.\footnote{The motivations for our modifications were to avoid numerical artifacts in the simulated density fields \citepalias{paper1}. These modifications are not required if one only wishes to model a given analytic bispectrum. Hence, the public CMB-BEST pipeline is adequate for the goals of this Appendix.} While there are differences in the analysis choices, both approaches provide accurate approximations to the templates under consideration, in the range they are being considered. Therefore, differences in the analysis choice are only a minor detail.

\section{Validation of Initial Conditions}\label{appx:Validation}

The transformations we perform on the Gaussian initial conditions can potentially lead to numerical artifacts, which will then bias the resolved structure in the $N$-body simulation. We confirm the absence of such issues in our models through the same tests used in \citetalias{paper1}. We direct the interested readier to Appendix B of that work for more details on the exact methods being employed. Here, we simply show the results and the associated discussions. All fields are generated on grids of $N_{\rm grid} = 128$ in a $L = 1 \gpc/h$ box, and we only show Fourier statistics up to half the Nyquist scale, $k_{\rm max} = 0.18 \hinvmpc$.

\begin{figure}
    \centering
    \includegraphics[width=\columnwidth]{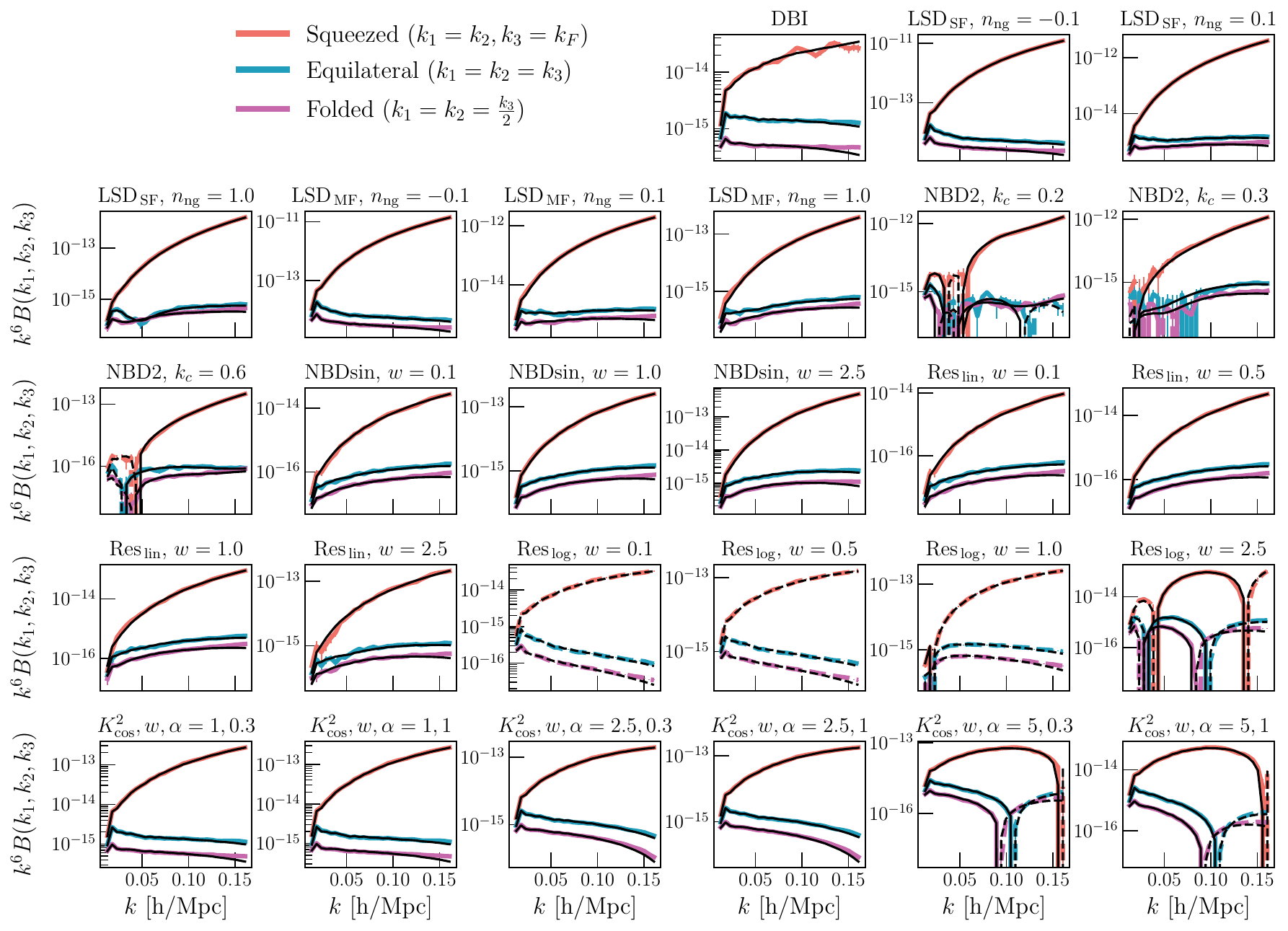}
    \caption{The average bispectrum (over 30 realizations) of the ICs compared to the theoretical expectation. We compute the measurement as the difference, $B(k) = (B^{\fNL = +X}(k) - B^{\fNL = -X}(k))/2$, which is noise-suppressed. The errorbars show the uncertainty on the mean bispectrum and are estimated using bootstrap realizations. In all cases, the measured bispectrum is in agreement with the theoretical expectation. We note a slight suppression in power in the folded limit at large wavenumbers. We have confirmed this effect is due to resolution and vanishes if we double the resolution of the grid used for the calculations.}
    \label{fig:Validation:Bispec}
\end{figure}

First, Figure \ref{fig:Validation:Bispec} shows that the bispectrum measured in the initial conditions matches the theoretical model used to modify the initial conditions and induce non-Gaussianity. There is a mild difference at large $k$ but we have confirmed this is due to the finite size of the grid used to resolve the initial density field. Increasing the grid resolution removes this feature, but increases the computational cost significantly, so we do not pursue this approach given the effect is only minor.

\begin{figure}
    \centering
    \includegraphics[width=\columnwidth]{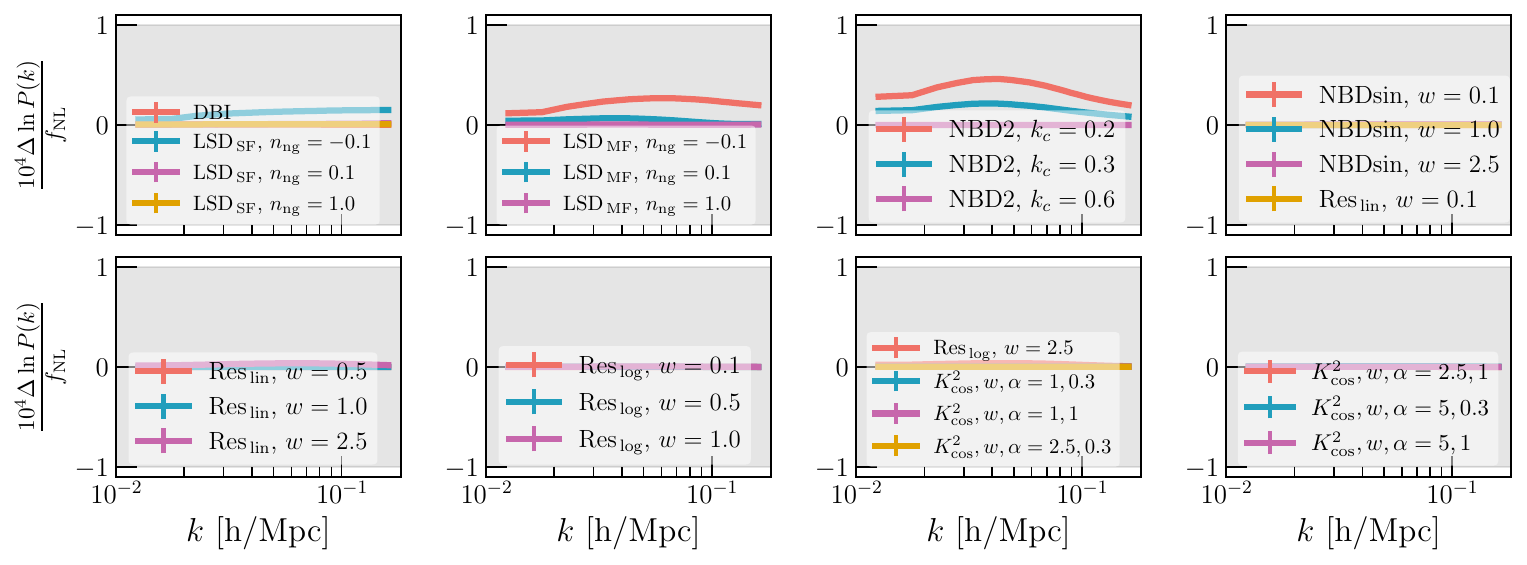}
    \caption{Measurements of the one-loop corrections (averaged over 30 realizations) to the power spectrum due to adding a given bispectrum model to the initial conditions. If the model has a relative change of $\Delta \ln P < 10^{-4}$ then a field with $\fNL = 100$ will have $<1\%$ correction to the power spectrum. We find all models are one to two orders of magnitude below this limit. Therefore, all corrections to the power spectrum are completely subdominant.}
    \label{fig:Validation:Pspec}
\end{figure}

Figure \ref{fig:Validation:Pspec} shows that the addition of a bispectrum to the initial conditions has a minimal impact on the primordial power spectrum. The fractional correction is well below $10^{-4}$ for unit change in $\fNL$, and therefore below 1\% for $\fNL = 100$. Following the arguments of \citetalias{paper1}, a 1\% change in the power spectrum is still below the current uncertainties on the primordial power spectrum amplitude \citep{Planck:2020:CosmoParams}. Thus, the measured changes to the power spectrum are completely negligible.

\begin{figure}
    \centering
    \includegraphics[width=\columnwidth]{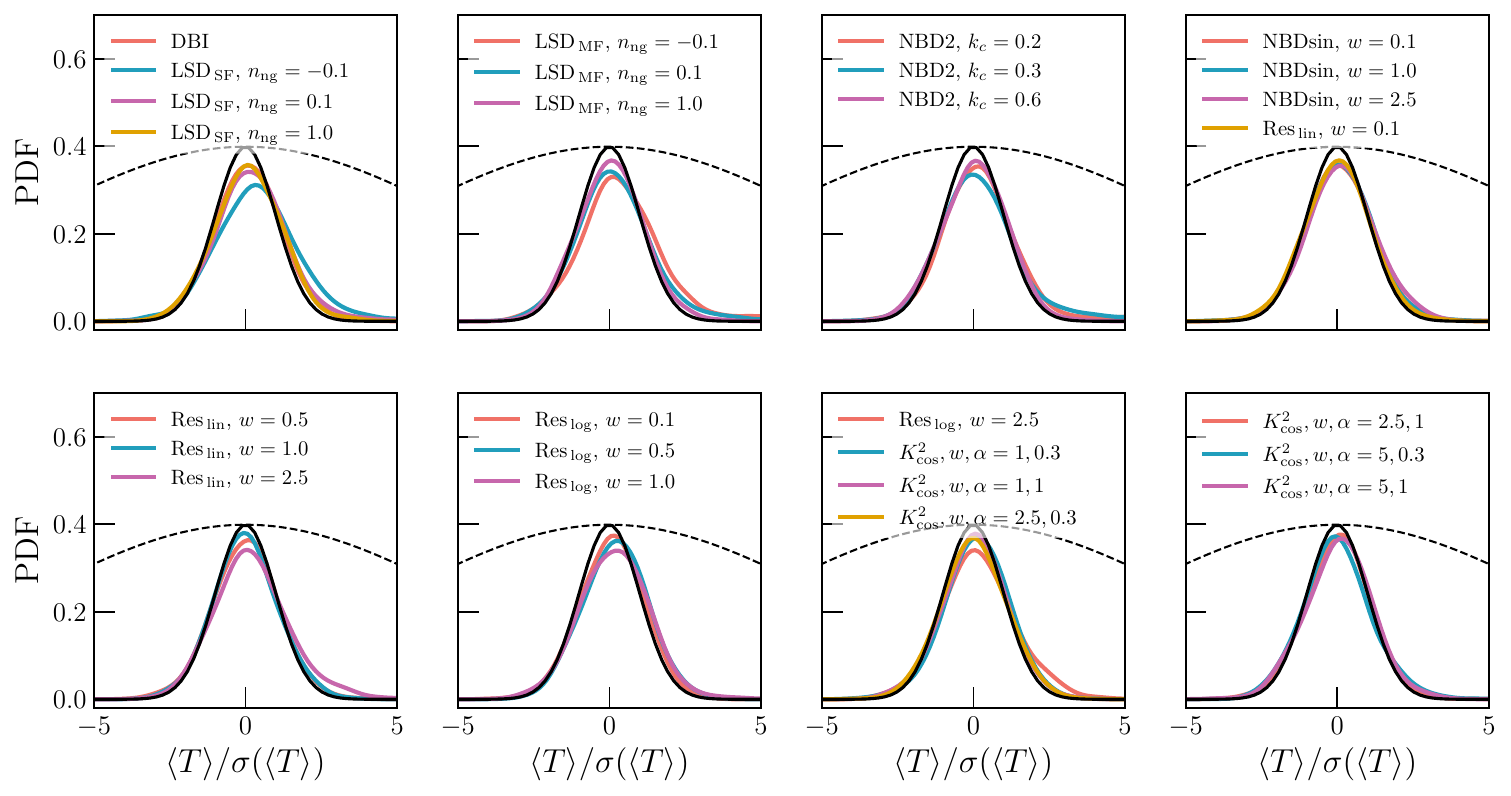}
    \caption{The measurement significance for the average trispectrum (measured over 30 realizations). We measure the significance of the mean trispectrum in each of the 2000 bins and present them as a histogram. A unit Gaussian is overplotted for reference. The measured significances closely match the Gaussian distribution, indicating the mean trispectrum is consistent with noise. We note that we are presenting the significance of the averaged, cosmic variance-suppressed measurement of the trispectrum. The above test is therefore a very conservative search for trispectrum signals in our ICs. The uncertainty for a single realization (not the mean) with no noise suppression is at least a factor of 7 larger (see \citetalias{paper1} for more details). We overplot a Gaussian with width $\sigma = 7$ as a dashed black line for reference.}
    \label{fig:Validation:Tk}
\end{figure}

Figure \ref{fig:Validation:Tk} confirms that our pipeline does not generate an unphysically large trispectrum in the initial conditions. The measured mean trispectrum is consistent with the null-signal for all models. The measured trispectrum is also subdominant to the noise level for a single simulation, and we show this noise with the dashed black line in each panel. Similar to \citetalias{paper1}, the uncertainty in measurements from a single simulation is taken to be a factor of $7\times$ larger than the uncertainties from the noise-suppressed, variance-suppressed trispectrum averaged over thirty realizations. In summary, the initial conditions do not show any notable trispectrum signal.

%%%%%%%%%%%%%%%%%%%%%%%%%%%%%%%%%%%%%%%%%%%%%%%%%%

% Don't change these lines
\label{lastpage}
\end{document}